\documentclass{aa}
\usepackage{graphicx}
\usepackage{graphics}
\usepackage{amssymb}                                                           
\usepackage{latexsym}
\begin{document}
\title{A new, cleaner colour-magnitude diagram for the metal-rich 
globular cluster NGC 6528 \thanks{Based on observations with the NASA/ESA 
Hubble Space Telescope, obtained at the Space Telescope Science 
Institute, which is operated by the Association of Universities 
for Research in Astronomy, Inc. under NASA contract No. NAS5-26555}}

   \subtitle{Velocity dispersion in the Bulge,  age and 
proper motion of NGC 6528}
   \titlerunning{A new colour-magnitude diagram for NGC 6528}

   \author{S. Feltzing
 \and R.A. Johnson
          }

   \offprints{S. Feltzing}

   \institute{Lund Observatory, Box 43, SE-221 00 Lund, Sweden, 
{\sl sofia@astro.lu.se}
         \and Institute of Astronomy, Madingley Road, CB3 0HA
         Cambridge, U.K., current address: 
	 ESO, Alonso de Cordova 3107, Santiago, Chile
         {\sl rjohnson@eso.org}
             }
   \date{Received 20/07/01 / Accepted 12/12/01}
 
\abstract{Using two epochs of HST/WFPC2 images of the metal-rich
globular cluster NGC 6528 we derive the proper motions of the  stars
and use them to separate the stars belonging to NGC 6528 from those of
the Galactic bulge. The stellar sequences in the  resulting
colour-magnitude diagram for the cluster are significantly better
determined than in previously published data. From comparison of the
colour-magnitude diagram with the fiducial line for NGC 6553  from
Zoccali et al.  (2001) we conclude that the two globular clusters have
the same age. Further, using $\alpha$-enhanced stellar isochrones, NGC
6528 is found to have an age of $11\pm 2$ Gyr. This fitting of
isochrones also give that the cluster is 7.2 kpc away from us. From
the measured velocities both the proper motion of the cluster and the
velocity dispersion in the Galactic bulge are found.  NGC 6528 is
found to have a proper motion relative to the Galactic bulge of
$<\mu_l>$ = 0.006 and $<\mu_{\rm b}>$ = 0.044 arcsec per century. Using
stars with $\sim 14 <V_{\rm 555} < 19$ (i.e. the red giant branch and
horizontal branch) we find, for the Galactic bulge,  $\sigma_{\rm l}=
0.33\pm 0.03$  and $\sigma_{b}= 0.25\pm0.02$ arcsec per century. This
give $\sigma_{\rm l}/\sigma_{\rm b}=1.32\pm0.16$,  consistent both with previous
proper motion studies of K giants in the  Galactic bulge as well as
with predictions by models of the kinematics of bulge stars.
}

\maketitle

%
%________________________________________________________________
 
\section{Introduction}

NGC 6528 is perhaps {\sl the} most metal-rich globular cluster known
and several studies have therefore targeted this cluster,
e.g. Ortolani et al. (1992), Ortolani et al. (1995), Richtler et
al. (1998), Cohen \& Sleeper (1995), Heitch \& Richtler (1999),
Carretta et al. (2001). It has also been used as a reference in
studies of other clusters, e.g.  Davidge (2000).   We summarize
important literature measurements of NGC6528 in Table
\ref{ngc6528.tab}.

NGC 6528 is at $(l,b)=(1.14,-4.12)$, i.e. in the plane of the Galactic
disk and towards the Galactic bulge.  The first effect of this is that
it is heavily reddened by foreground dust, Ortolani et al. (1992),
Richtler et al. (1998), Heitsch \& Richtler (1999). Most recent
distance estimates put the cluster within less than 1 kpc from the
Galactic centre (e.g. Richtler et al. 1998 and references
therein). The close proximity to bulge stars further complicates the
interpretation of the colour-magnitude diagram. Since the bulge stars
and the cluster stars have roughly the same distance modulus they are
superimposed in the colour-magnitude diagram. This effect has been
noted to be particularly pronounced in the red giant branch (Richtler
et al. 1998).

\begin{table*}
\caption[]{Data for NGC6528 compiled from the literature.}
\label{ngc6528.tab}
\begin{tabular}{llllllllllllllllll}
\hline\noalign{\smallskip}
  & Value & Ref. & Comment\\\noalign{\smallskip}
\hline\noalign{\smallskip}
Core radius & $0\farcm 09$& Harris (1996) \\
 Distance & 7.5 kpc  & Ortolani et al.\,(1992) \\
$\Delta(m-M)$ & 16.4 & Zinn (1980) & 19.1 kpc\\
              & 14.35 &  Ferraro et al. (1999) \\
              & 15.15$\pm$0.24&  Heitsch \& Richtler (1999) & isochrone fitting \\
 {[Fe/H]}     & $+0.01$ & Zinn (1980)\\
              & $+0.29$ & Bica \& Patoriza (1983)\\ 
              & $+0.12$ & Zinn \& West (1984) \\
              & $-0.23$ & Armandroff \& Zinn (1988) & Integrated 
spectra IR Ca {
\sc ii}\\
              & high, sim to NGC6553& Ortolani et al.\,(1992)\\
              & $-0.23$ &Origlia et al.\,(1997) & IR abs. at 1.6 $\mu$m \\
              &  $+0.07\pm0.1?$ & Carretta et a. (2001) \\
 {[M/H]} & $+0.1/-0.4$ & Richtler et al.\,(1998)&Trippico isochrone/Bertelli isochrone\\
         & -0.31 &  Ferraro et al. (1999) \\
         & 0.00 & Heitsch \& Richtler (1999)          \\
   $Z$ & $Z_{\odot}$ & Bruzual et al.\,(1997) & \\
 Age & 14 Gyr & Ortolani et al.\,(1992) & metallicity comparable to solar\\
        & $12\pm2$ Gyr & Bruzual et al.\,(1997) & \\
 E(B-V) &  0.56 & Zinn (1980) \\
        &  0.55 & Ortolani et al.\,(1992) & NGC6553 as reference and
$\Delta(m-M)_V= 14.39$\\
        & 0.62 & Bruzual et al.\,(1997) & \\
        & 0.62 & Ferraro et al. (1999) \\
  E(V-I)& 0.8/0.6 & Richtler et al.\,(1998)&Trippico isochrone/Bertelli isochrone\\
        & 0.46$\pm$0.03&   Heitsch \& Richtler (1999)           & isochrone fitting\\
\noalign{\smallskip}
\hline\noalign{\smallskip}
\end{tabular}
\end{table*}

In a pre-study we noted in particular that the colour-magnitude
diagram of the globular cluster NGC 6528 closely resembled that of NGC
5927 {\sl if} the latter was superimposed on the colour-magnitude
diagram of Baade's window (for a set of representative
colour-magnitude diagrams see Feltzing \& Gilmore 2000).  This is
consistent with the cluster being virtually inside the Galactic bulge
and thus having a large contribution of bulge stars in its
colour-magnitude diagram. Moreover, if the cluster is as metal-rich as
indicated in previous studies (and now confirmed by Carretta et
al. 2001) then the red-giant branch as well as both the turn-off and
the horizontal branch of the bulge and globular cluster will appear at
virtually the same magnitudes and colours. The bulge stars will be
more spread out in the colour-magnitude diagram than those in the
cluster, due to the large range of metallicities and ages present in
the bulge (e.g.  McWilliam \& Rich 1994, Feltzing \& Gilmore 2000).
Thus the only way to obtain a clean colour-magnitude diagram for the
cluster is to obtain proper motions of the cluster stars relative to
the bulge stars and separate the two populations using their proper
motions.

This conclusion prompted us to apply for HST time to obtain a  second
epoch of observations of NGC 6528 with WFPC2 in order to derive the
relative proper motion of the cluster as compared to that of the
Galactic bulge.

We report here on the results from this proper motion study. The
article is organized as follows; Sect. 2 presents the data, derivation
of photometry and proper motions are discussed in Sect. 3 and 4,
Sect. 5 and 6.  contains an extensive discussion of mean proper
motions and velocity dispersions for  NGC 6528 relative  to the
Galactic bulge as well as the velocity dispersion in the bulge itself,
in Sect. 7 we use the proper motion information to obtain  a clean
colour-magnitude diagram for NGC 6528, Sect. 8 contains  a discussion
of the age for the cluster, and finally Sect. 9  provides a brief
discussion and conclusions.

\section{The data}

The data consist of two sets of observations, one from the HST archive
and observed in 1994 (GO 5436) and the other our new data for the same
field (GO 8696, PI Feltzing). We detail the number of exposures in
each filter and epoch as well as  integration times in Table
\ref{data.tab}. The first data set was taken on the 27 February 1994
and the new one (GO 8696) on the 1 April 2000. This gives a time  span
of 6.093 years between the two set of observations.

\begin{table}
\caption{NGC6528 HST/WFPC2 observations used in this study}
\begin{tabular}{llll}
\hline\noalign{\smallskip}
GO & Filter & \multicolumn{2}{c}{ Exp. time (s)}\\
\noalign{\smallskip}
\hline\noalign{\smallskip}
5436 & F555W & $2\times 100$  & $1\times 5$ \\
5436 & F814W & $2 \times 50$  & $1\times 14$ \\
8696 & F555W & $8 \times 100$ & $1\times 5 $ \\
8696 & F814W & $2 \times 50$ \\ 
\hline\noalign{\smallskip}
\end{tabular}
\label{data.tab}
\end{table}

The first epoch of observations provided a long and a short set  of
exposures in F555W and F814W with the cluster centre on WF3.  Our new
observations were observed with the same WFPC2 orientation, but with
four times as long total exposure time in F555W.  The increased
exposure times enable us to reliably detect all the possible stellar
sources in the first epoch data.

A new pair of F814W images were also taken in order to improve the
accuracy of the F814W magnitudes.  Finally, one short exposure in
F555W was also obtained in order to find the proper motions of the
brightest stars in the field which are saturated in the longest
exposures.

The images for each filter/epoch combination were combined using  the
{\sc drizzle} and {\sc crrej} tasks in the {\sc stsdas} environment
within {\sc iraf}\footnote{IRAF  is distributed by National Optical
Astronomy Observatories, operated by the Association of Universities
for Research in Astronomy, Inc., under contract with the National
Science Foundation, USA.}.  Drizzling removes the WFPC2 geometric
distortion from the images.

For the first epoch data we combined the F555W and F814W images into
one single image to obtain as deep an image as possible for object
centering.

\section{Stellar photometry}
\label{sect:stellarphot}

Stellar photometry was done inside the {\sc digiphot.daophot}  package
in {\sc iraf}. In particular we found the stars using {\sc daofind} on
the deep, new F555W images, then aperture and psf-photometry were
obtained using {\sc phot} and {\sc allstar} on the same images. For
the aperture photometry we used an aperture of 2 pixels.  For the
psf-photometry psfs variable over the chips, constructed from the
images themselves, were used and the output statistics, i.e. $\chi$
and sharpness, were used to weed out non-stellar sources.

This created a list of stars that were then used for obtaining
aperture photometry from the new F814W images, which together with the
F555W data gave a first colour-magnitude diagram.  The positions of
the stars making up the colour-magnitude diagram provided our source
list of objects to find proper motions for.

For the images with short exposures  we performed aperture photometry
on  the first epoch   images for each filter and then merged the F555W
and F814W data into a colour-magnitude diagram. Keeping only stars
detected in both F555W and F814W this provided a  master list for
bright objects to find the proper motions for in the  short exposures.

The final colour-magnitude diagrams are all based on the aperture
photometry.

\subsection{Calibration}
\label{sect:calib}

We obtained instrumental ($-2.5\log(counts/s)$,) magnitudes from both
the old, short and new, long exposure data. These magnitudes were
corrected to the HST/WFPC2 magnitude system, and merged to form a
final single photometry set as described below.

The final stellar magnitudes within the HST/WFPC2 system
i.e. V$_{555}$ and I$_{814}$, were obtained by applying a number of
corrections to the instrumental magnitudes, essentially following
Holtzmann et al. (1995b).

The photometry was obtained from drizzled images. The drizzling
procedure removes the geometric distortion from the images, therefore
no geometric distortion correction was required for the photometry.

Aperture corrections to a 0.5 arcsec aperture  were obtained from our
own images.  These corrections were allowed to vary with distance from
the centre of the chip i.e. correction $= a + b\times$distance The
values of the constants $a$ and $b$ are given in Table
\ref{tab:apcorr}.  For the WF4 long data there were not enough stars
to obtain both $a$ and $b$ from our data. In this case we used the
value of $b$ from Gonzaga et al. (1999) and only fit the value of $a$
from our data.

\begin{table}
\caption{Values of constants found for aperture correction in each image,
see Sect. \ref{sect:calib}}
\begin{tabular}{lllll}
\hline\noalign{\smallskip}
 & \multicolumn{2}{c}{F555W} & \multicolumn{2}{c}{F814W} \\
    &\multicolumn{1}{c}{$a$}&\multicolumn{1}{c}{$b$}&\multicolumn{1}{c}{$a$}&
\multicolumn{1}{c}{$b$}\\
\noalign{\smallskip}
\hline\noalign{\smallskip}
\multicolumn{5}{l}{Short} \\
\noalign{\smallskip}
\hline\noalign{\smallskip}
WF2 & 0.17 & 1.70e-4 & 0.21 & 1.78e-4 \\
WF3 & 0.21 & 1.47e-4 & 0.23 & 1.90e-4 \\
WF4 & 0.27 & 2.21e-4 & 0.23 & 1.59e-4 \\
\hline\noalign{\smallskip}
\multicolumn{5}{l}{Long} \\
\noalign{\smallskip}
\hline\noalign{\smallskip}
WF2 & 0.24 & 5.42e-5 & 0.27 & 4.35e-5 \\ 
WF3 & 0.24 & 2.16e-4 & 0.28 & 1.84e-4 \\
WF4 & 0.24 & 1.17e-4 & 0.23 & 1.07e-4 \\
\noalign{\smallskip}
\hline\noalign{\smallskip}
\end{tabular}
\label{tab:apcorr}
\end{table}

Corrections for the charge transfer (in)efficiency (CTE) were
applied. The short data were obtained before the WFPC2 cool-down on
23/4/1994. The CTE problem is worse for the pre-cool-down data, but the
corrections are not well modeled. For these data the CTE problem was
corrected by assuming that the star counts lost were a linear function
of the $y$-position on the chip, with stars at the top of the chip
losing the maximum 10\% (Holtzmann,1995b). For our new long data we
used the equations in Whitmore et al. (1999) to correct for CTE.

The zero-points used to transform to HST/WFPC2
 magnitudes are from Baggett et al. (1997).

\subsection{Merging long and short photometry}

To make the final colour magnitude diagrams we have merged the
photometry from the long and short exposures. We compared the
magnitudes of the same stars from the short and long exposure data,
and found no evidence for any offsets. Further, we found that the long
exposure data were saturated for V$_{555}<$16.8 and I$_{814}<$15.

The two sets of long and short photometry were then merged according
to the following rules
\begin{itemize}
\item  if the star has both long and short photometry then 
use long if V$_{555}>$16.8 and I$_{814}>$15, otherwise use short
\item if the star just has short photometry then use it if 
V$_{555}\le$16.8 or I$_{814}\le$15
\item if the star just has long photometry then use it if 
V$_{555}>$16.8 and I$_{814}>$15
\end{itemize}

\begin{figure}
%\resizebox{16cm}{!}{\includegraphics[width=16cm]{h3031f01.ps}}
\resizebox{10.5cm}{!}{\includegraphics[width=16cm]{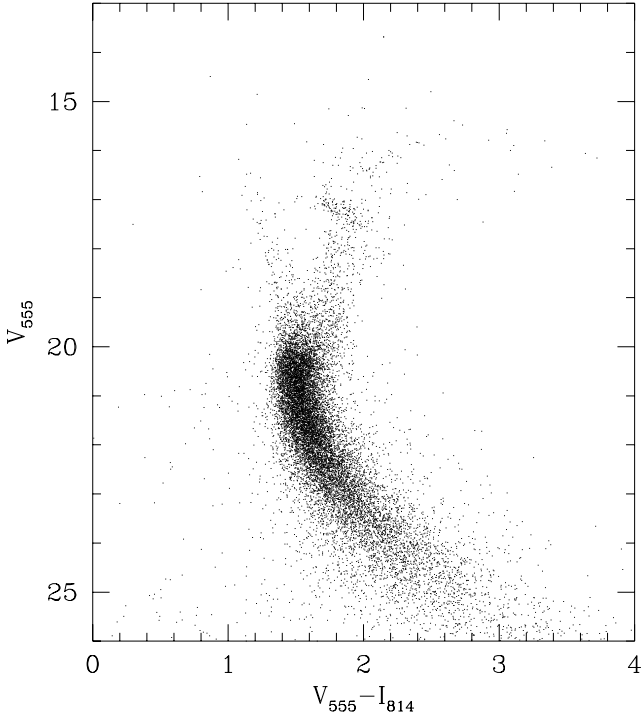}}
\caption[]{Colour-magnitude diagrams for all three WF. 
All stars measured in both the new and old
images and which satisfied the cuts imposed in $\chi$ and sharpness
variables (see Sect. \ref{sect:stellarphot}) on the individual WFs are included.
Stars with fitting errors are excluded }
\label{cmdall.fig}
\end{figure}

Our final colour-magnitude diagram is shown in Fig. \ref{cmdall.fig}.
This includes all stars selected according to our selection criteria
for $\chi$ and sharpness. We also require the stars to all have  good
positions according to the photometric routines in {\sc daophot}.  No
corrections for differential reddening have been applied.

\section{Proper motions}

The heart of this investigation is to use the measured proper motions
to separate out the bulge and cluster stars.  Our chosen method to
find the proper motion for each star consists of the following steps:

\begin{enumerate}

\item find the transformation, $x$- and $y$-shifts as well as rotation (small
in our case), between the two epochs of observations using a small
number of bright stars

\item transform the positions on the new images (the deep drizzled
F555W images in our case) to the reference frame of the old images

\item re-center the transformed positions on the old image (in our
case we used the deep image combined from both filters for this)

\item inspect the $x$- and $y$-shifts found in the re-centering, select
only stars that have small shifts relative to the centre of the 
distribution of shifts

\item using only those stars selected in the previous item
find a new, improved transformation between the epochs

\item apply the new solution and re-center the stars on the old image

\item if deemed necessary iterate 

\end{enumerate}

We only needed to iterate a few times to find our solutions. Since the
WF chips in the WFPC2 over time slowly drift relative to each other,
Fruchter \&  Mutchler (1998), the transformations between the new and
the old reference frames  have to be found separately for each
WF-chip.

 For the short exposures we first applied the transformation found
from the deep images. However, since it appears that the new short
exposure is not perfectly aligned with the new deep images we iterated
the solution for the transformation once.
 
\section{Measuring proper motions and velocity dispersions}

\begin{figure}
\resizebox{\hsize}{!}{\includegraphics{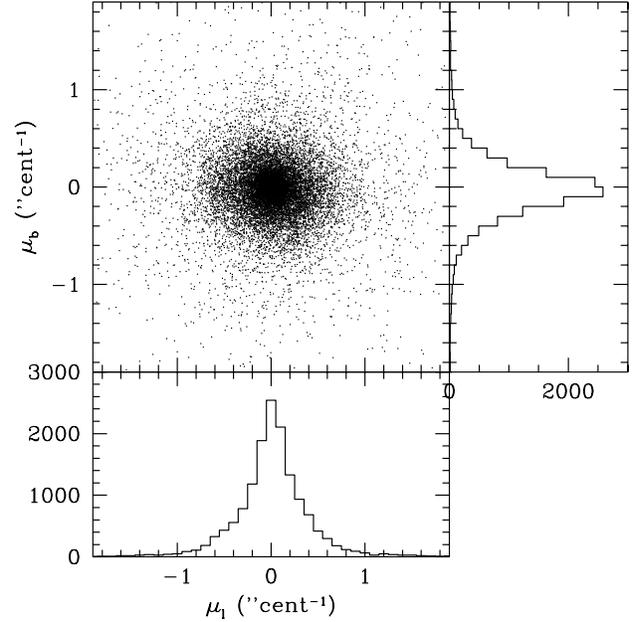}}
\caption[]{Stellar proper motions and the resulting histograms
for $\mu_{\rm l}$ and $\mu_{\rm b}$. The proper motions are measured 
in arcsec per century }
\label{shiftshist.fig}
\end{figure}

The shifts obtained (in pixels) for the individual  WFs were
transformed to  a common grid, i.e. the Galactic coordinates, as well
as to  arcsec per century ($''$cent$^{-1}$).  In
Fig. \ref{shiftshist.fig} these proper motions are plotted together
with the histograms for the proper motions in $l$ and $b$. The full
data set for stars brighter than $V_{\rm 555}=19$ (proper motions,
positions, and magnitudes) are given in the Tables in Appendix
\ref{app:prop}.

In an ideal scenario, where the globular cluster has an appreciable
motion in relation to the bulge stellar population, the bulge and
cluster stars will form two distinct distributions in the proper
motion diagram.  See for example the recent results by Zoccali et
al. (2001) for NGC 6553, King et al. (1998) for NGC 6739 and Bedin et
al. (2001) for M4, or those for NGC 6712 by Cudworth (1988).  The
bulge stars have a larger velocity dispersion than the cluster stars,
and hence a larger scatter in the proper motion diagram.

In our case Fig. \ref{shiftshist.fig} shows that the motion of the
globular cluster in relation to the Galactic bulge is very small.
This is as expected since the heliocentric radial velocity  relative
to the local standard of rest for NGC 6528 is high ($184.9\pm3.8$ km
s$^{-1}$ (Harris 1996), $\simeq$210 km s$^{-1}$ (Carretta et
al. 2001)), which suggests that NGC6528 is on a mostly radial orbit
away from us.  However, we note that the histograms for the velocities
in the $l$ and $b$ coordinates have broad wings.  In fact when we
tried to fit our histograms with Gaussian distributions it became
clear that a single Gaussian distribution could not fit the observed
distributions and two Gaussians were needed.

\begin{figure*}
\resizebox{15cm}{!}{\includegraphics{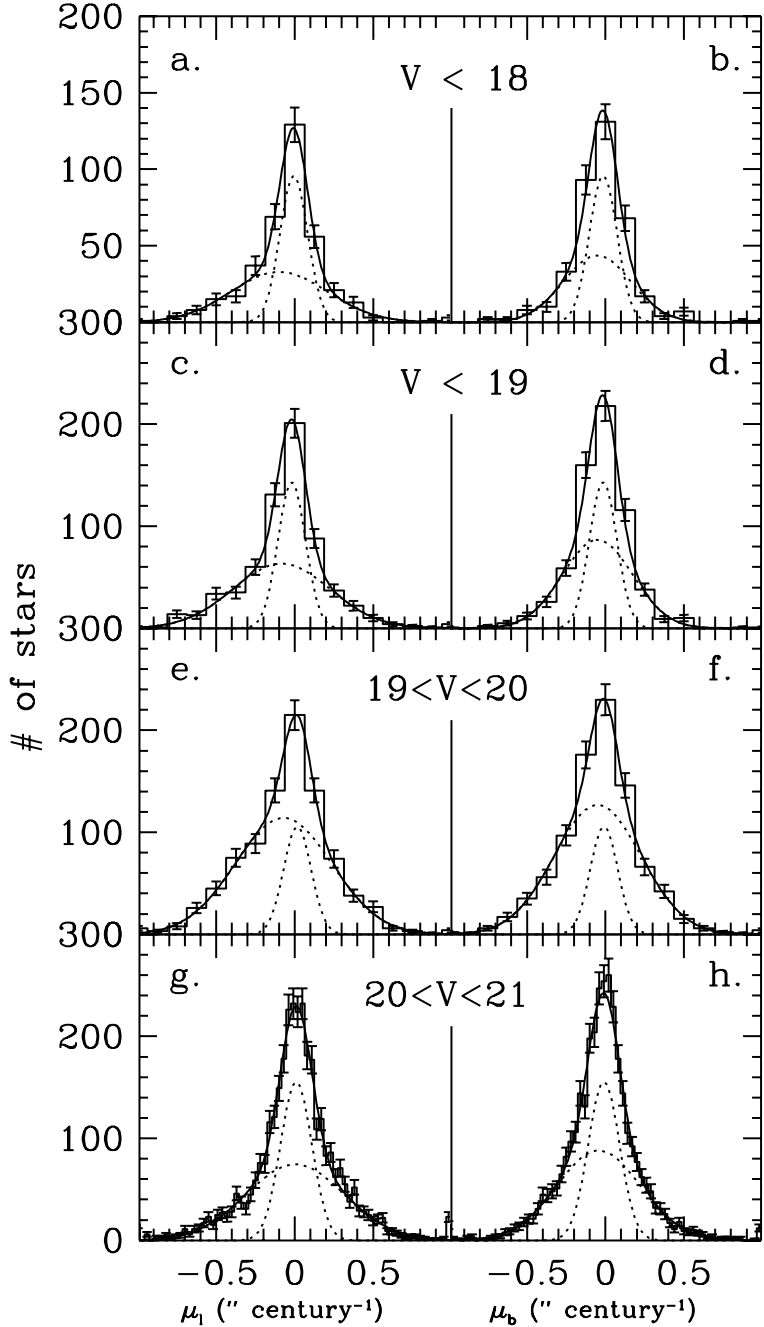}}
\caption[]{Histograms for the $l$ and $b$ proper motions ($\mu_{\rm l}$, $\mu_{\rm b}$)
for different magnitude ranges (as indicated). For each histogram Poissonian
errorbars are drawn as well as the fitted Gaussians. The dotted lines show the
two Gaussians needed to fit the data and the full line the final combined
distribution.
The fitted parameters are given in Table \ref{hist_all.tab}.
 Proper motions are measured in arcsec per century }
\label{hist_all.fig}
\end{figure*}

To separate the bulge and cluster stars using the measured proper
motions, we divide the stars into different magnitude ranges and find
the best fitting Gaussians, as shown in Fig.  \ref{hist_all.fig}.  We
found that two Gaussians were required to fit the data well,
indicating that, as expected, we have two stellar populations with
different velocity dispersions.  Based on previous measurements of
bulge and cluster velocity dispersions, we associate the narrow
Gaussian with NGC 6528 and the broad Gaussian with the bulge stars.

We found that the $\mu_{\rm l}$-diagram was easily fit by two
Gaussians by our routine, however $\mu_{\rm b}$ proved more difficult.  In
fact the fitting routine found one badly fitting broad Gaussian for
the data in Fig. \ref{hist_all.fig}. 
However we expect the cluster stars to have the same
velocity dispersion and amplitude in $\sigma_{\rm l}$ and $\sigma_{\rm b}$ and so
we fixed the parameters for the narrower Gaussian, which represents
the cluster stars, from the fitting of the $\mu_{\rm l}$-distribution.  In
particular we fixed the height and the width but left the position
free to be fitted.  The second Gaussian had all three parameters
(width, height and position) free for fitting.  The results for all
the fitted Gaussians are give in Table \ref{hist_all.tab}.

\begin{table*}
\caption{Gaussian fits to proper motion distributions in Fig. 
\ref{hist_all.fig}. An error of 0. indicates that that particular
parameter was kept fixed during the fitting procedure. We give
the amplitude, the center ($<\mu>$), and the 
$\sigma$ for two Gaussians (shown with dotted lines in Fig. 
\ref{hist_all.fig}). Both are in arcsec per century. 
The first fours rows give the results for 
the $l$-coordinate and the last for for the $b$-coordinate }
\begin{tabular}{lrrrrrr}
\hline\noalign{\smallskip}
\multicolumn{1}{l}{Galactic $l$} & \multicolumn{3}{c}{Field} & \multicolumn{3}{c}{Cluster}\\
\noalign{\smallskip}
\hline\noalign{\smallskip}
\multicolumn{1}{c}{mag. range}&\multicolumn{1}{c}{$Amp_1$}&\multicolumn{1}{c}{$<\mu_1>$}&\multicolumn{1}{c}{$\sigma_1$}&\multicolumn{1}{c}{$Amp_2$}&\multicolumn{1}{c}{$<\mu_2>$}&\multicolumn{1}{c}{$\sigma_2$} \\
 & & \multicolumn{2}{c}{(arcsec century$^{-1}$)} & & \multicolumn{2}{c}{(arcsec century $^{-1}$) } \\
\noalign{\smallskip}
\hline\noalign{\smallskip}
V$<$18        & 32.783$\pm$5.990& --0.097$\pm$0.027&0.305$\pm$0.025 &95.539$\pm$ 11.934& --0.005$\pm$0.010& 0.088$\pm$0.011\\
V$<$19        & 63.362$\pm$7.231& --0.084$\pm$0.017&0.306$\pm$0.012 &142.831$\pm$14.992& --0.018$\pm$0.009& 0.088$\pm$0.009\\
19$\le$V$<$20 &114.083$\pm$9.793& --0.072$\pm$0.013&0.305$\pm$0.009 &105.318$\pm$16.496&   0.017$\pm$0.014& 0.087$\pm$0.015 \\
20$\le$V$<$21 & 74.461$\pm$4.957&   0.004$\pm$0.008&0.309$\pm$0.007 &155.211$\pm$ 7.769&   0.010$\pm$0.005& 0.099$\pm$0.006\\
\noalign{\smallskip}
\hline\noalign{\smallskip}
\multicolumn{1}{l}{Galactic $b$} & \multicolumn{3}{c}{Field} & \multicolumn{3}{c}{Cluster}\\
\noalign{\smallskip}
\hline\noalign{\smallskip}
V$<$18       &  43.627$\pm$5.509& --0.052$\pm$0.018& 0.214$\pm$0.014&  95.539$\pm$0.& --0.014$\pm$0.012& 0.088$\pm$0.\\
V$<$19       &  86.681$\pm$7.088& --0.051$\pm$0.013& 0.221$\pm$0.009& 142.831$\pm$0.& --0.015$\pm$0.010& 0.088$\pm$0.\\
19$\le$V$<$20& 126.907$\pm$6.433& --0.050$\pm$0.011& 0.273$\pm$0.007& 105.318$\pm$0.& --0.009$\pm$0.015& 0.088$\pm$0.\\
20$\le$V$<$21&  87.990$\pm$3.073& --0.044$\pm$0.007& 0.265$\pm$0.004& 155.211$\pm$0.& --0.010$\pm$0.005& 0.099$\pm$0.\\
\noalign{\smallskip}
\hline
\end{tabular}
\label{hist_all.tab}
\end{table*}

We find that all four $\mu_{\rm l}$-distributions are fit by two Gaussians,
one narrow and with a $\sigma_{\rm l} \approx 0.08$ arcsec per century, and
one broader with $\sigma_{\rm l} \approx 0.30$ arcsec per century. The
centers of these Gaussians vary with magnitude range, see Table
\ref{hist_all.tab}. For the broad Gaussian in $\mu_{\rm l}$ the three
brightest magnitude ranges agree very well within the calculated
errors while for the last bin the center has moved from $\sim -0.07$
to 0 arcsec per century. For the narrow Gaussian the centers for the
two brightest magnitudes agree within the errors, as do those for the
two faintest magnitudes.

For $\mu_{\rm b}$ we fixed the fwhm and height for the narrow Gaussian
before fitting (see discussion above). Thus by definition
$\sigma_{\rm b}=\sigma_{\rm l}$ and $amplitude_b=amplitude_l$ for each magnitude
range for the narrow Gaussian. For the broad Gaussian we find a
$\sigma_{\rm b}$ around $0.21$.  Unlike the $\mu_{\rm l}$ distribution, the fitted
centres for the broad and narrow Gaussians in $\mu_{\rm b}$ remain
consistent within the errors for all the magnitude bins.  The centers
for the two Gaussians are found to be $-0.05 \pm 0.01$ and $-0.012\pm
0.012$ arcsec per century.

The different behaviour of the centers of the $\mu_{\rm l}$ and $\mu_{\rm b}$
histograms with magnitude suggests the presence of a stellar component
in addition to the bulge and cluster stars that is adding an
additional proper motion component to $\mu_{\rm l}$. One possibility for
this component is disk stars, which may be the main component of the
blue plume seen in the colour-magnitude diagram in Figure
\ref{cmdall.fig}.  We therefore investigate below whether we see any
difference in Gaussians fit separately to blue and red samples.

\begin{figure*}
\resizebox{\hsize}{!}{\includegraphics{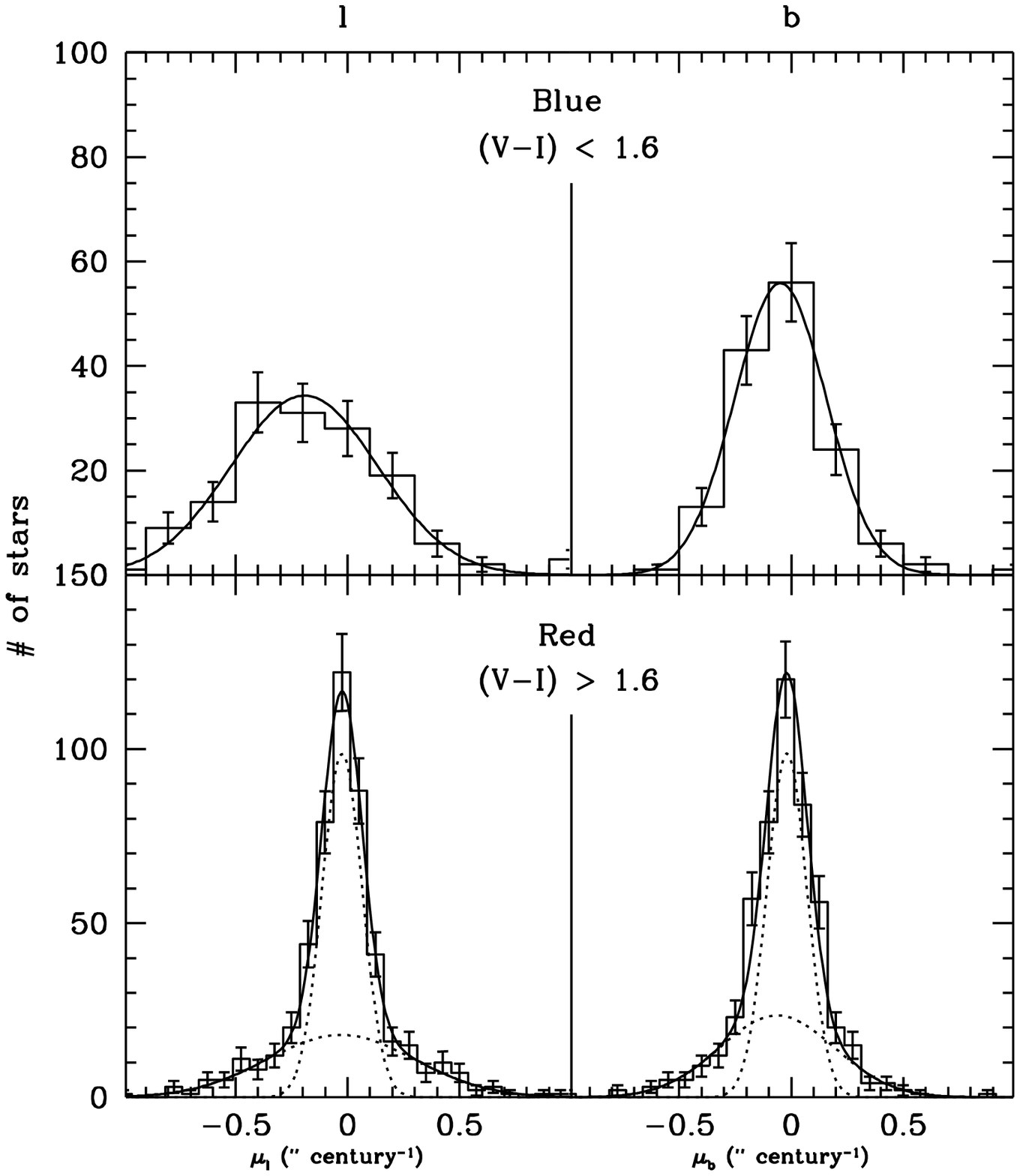}
\includegraphics{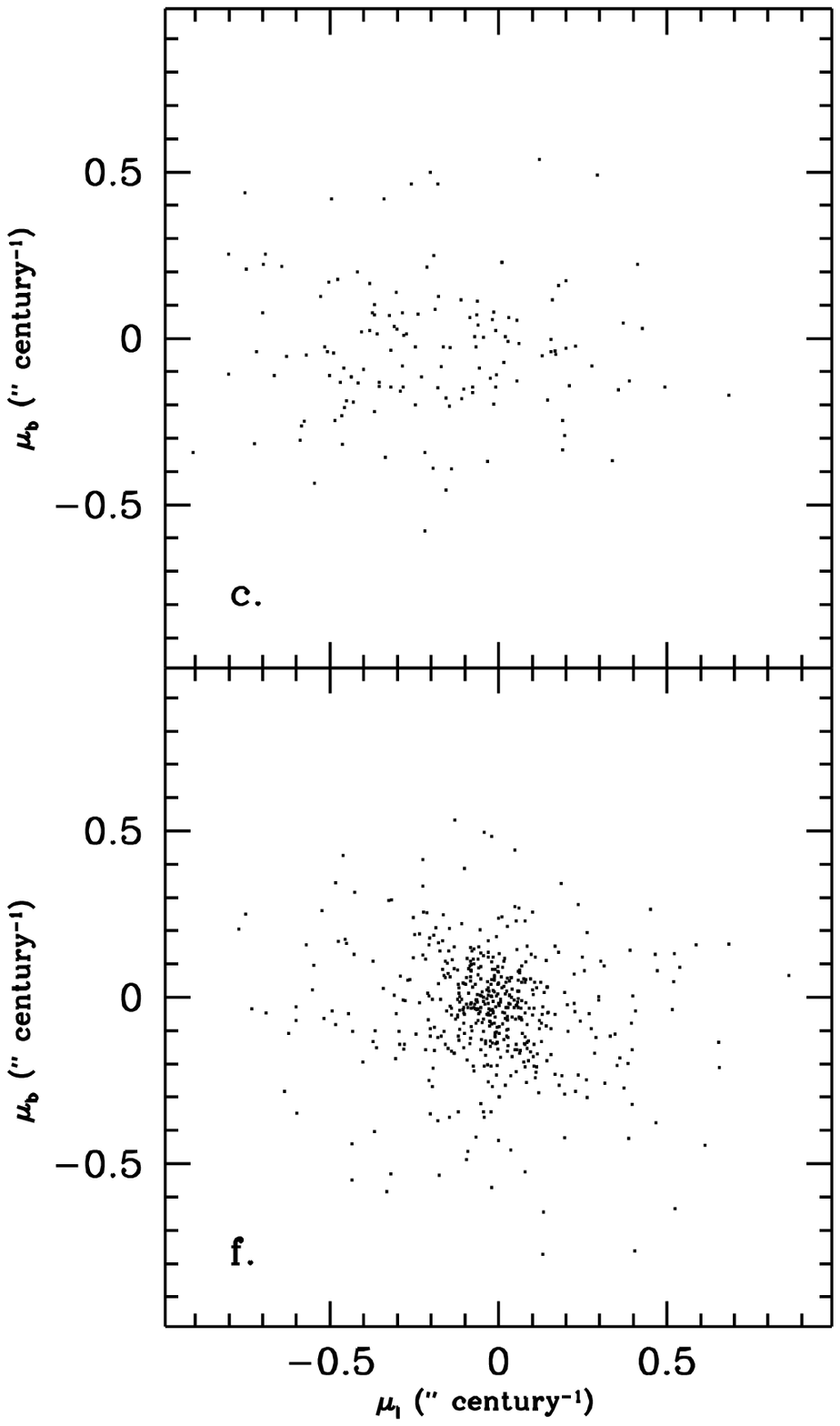}}
\caption[]{Histograms for the $b$ and $l$ proper motions ($\mu_{\rm l}$,
$\mu_{\rm b}$) for blue, $V-I<1.6$, and red, $V-I>1.6$, stars all with
$V_{\rm 555}\leq19$.  For each histogram Possonian errorbars are drawn
as well as the fitted Gaussians.  For the blue samples only one
Gaussian is fitted, while for the red two Gaussians are necessary to
fit the data (see text for discussion).  The fitted parameters are
given in Table \ref{hist_redblue.tab}.  Proper motions are measured in
arcsec per century }
\label{hist_redblue.fig}
\end{figure*}

\begin{figure*}
\resizebox{\hsize}{!}{\includegraphics{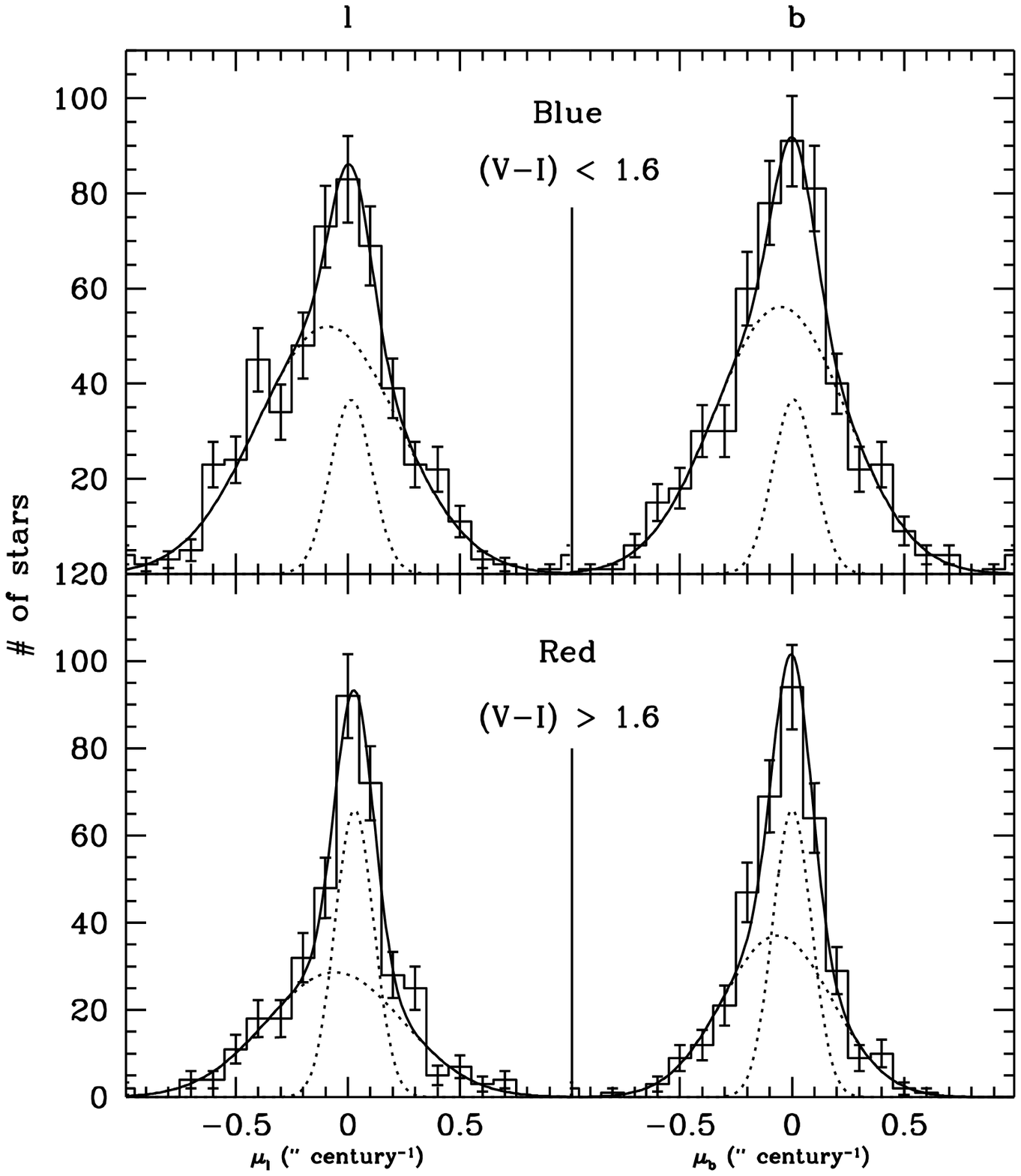}
\includegraphics{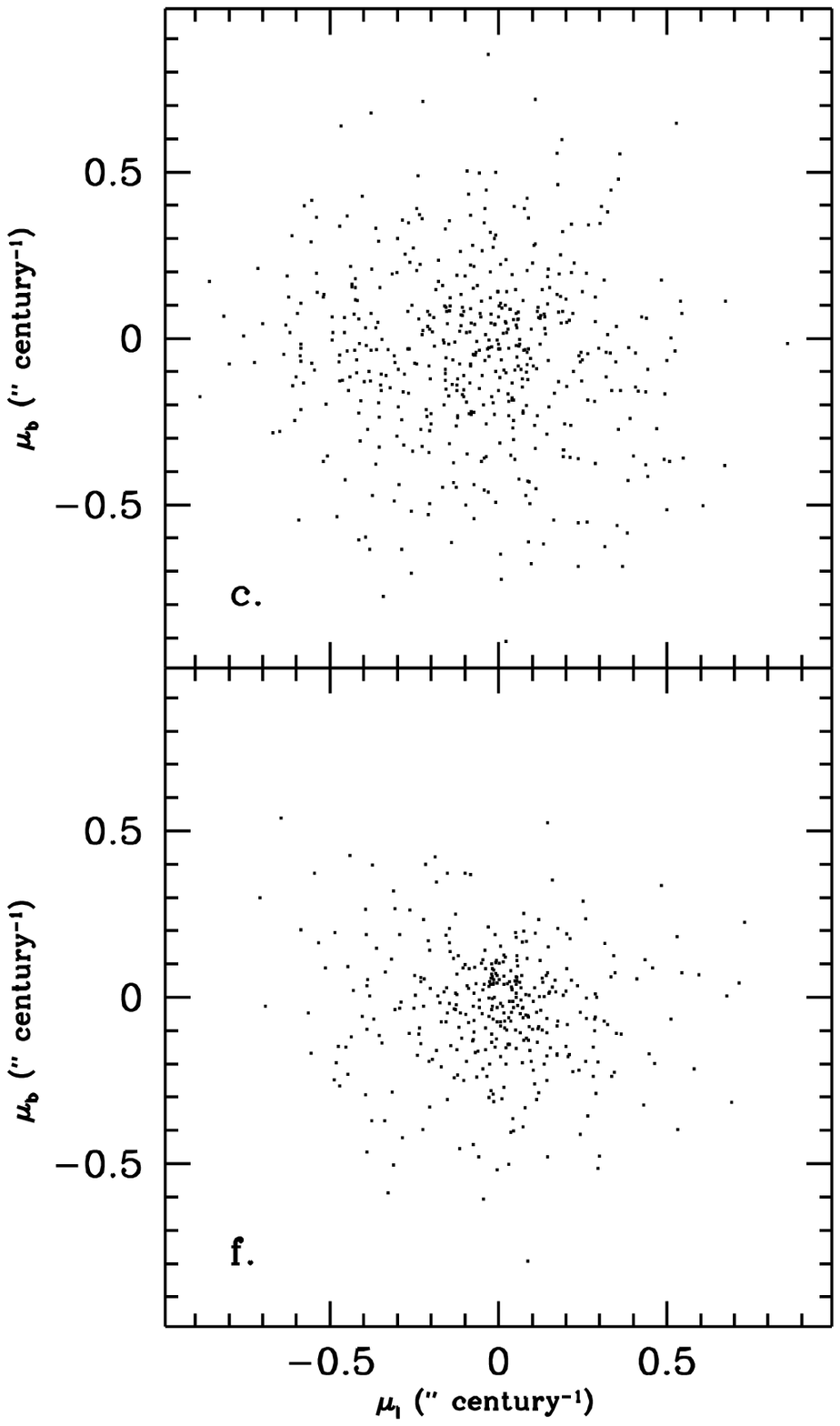}}
\caption[]{Histograms for the $b$ and $l$ proper motions ($\mu_{\rm l}$,
$\mu_{\rm b}$) for blue, $V-I<1.6$, and red, $V-I>1.6$, stars all with $19 <
V_{\rm 555}<20$.  For each histogram Possonian errorbars are drawn as
well as the fitted Gaussians.  For the blue samples only one Gaussian
is fitted, while for the red two Gaussians are necessary to fit the
data (see text for discussion).  The fitted parameters are given in
Table \ref{hist_redblue.tab}.  Proper motions are measured in arcsec
per century }
\label{hist_redblue_v1920.fig}
\end{figure*}

In Fig. \ref{hist_redblue.fig} and \ref{hist_redblue_v1920.fig} we
divide the two bright magnitude samples, $V_{\rm 555}\leq19$ and $19 <
V_{\rm 555}<20$ into red and blue stars and fit Gaussians to them in
the same way as before.  The fitted parameters are given in Table
\ref{hist_redblue.tab}.

\begin{table*}
\caption{Gaussian fits to proper motion distribution in
Fig. \ref{hist_redblue.fig} and Fig. \ref{hist_redblue_v1920.fig}. An
error of 0. indicates that that particular parameter was kept fixed
during the fitting procedure. We give the amplitude, the center
($<\mu>$), and the $\sigma$ for two Gaussians (shown with dotted lines
in the figures).  Both are in arcsec per century. }
\begin{tabular}{lrrrrrrr}
\hline\noalign{\smallskip}
\multicolumn{1}{l}{Galactic $l$} && \multicolumn{3}{c}{Field} & \multicolumn{3}{c}{Cluster}\\
\noalign{\smallskip}
\hline\noalign{\smallskip}
\multicolumn{1}{c}{Mag. range}&\multicolumn{1}{c}{Colour}&\multicolumn{1}{c}{$Amp_1$}&\multicolumn{1}{c}{$<\mu_1>$}&\multicolumn{1}{c}{$\sigma_1$}&\multicolumn{1}{c}{$Amp_2$}&\multicolumn{1}{c}{$<\mu_2>$}&\multicolumn{1}{c}{$\sigma_2$} \\
 & & &\multicolumn{2}{c}{(arcsec century$^{-1}$)} & & \multicolumn{2}{c}{(arcsec century $^{-1}$) } \\
\noalign{\smallskip}
\hline\noalign{\smallskip}
V$<19$      &$V-I<1.6$& 34.332$\pm$3.485& --0.193$\pm$0.028& 0.324$\pm$0.020& & & \\
V$<19$      &$V-I>1.6$& 17.912$\pm$3.496& --0.030$\pm$0.026& 0.327$\pm$0.027& 98.707$\pm$8.215& --0.024$\pm$0.007& 0.089$\pm$0.007\\
$19<$V$<20$ &$V-I<1.6$& 51.985$\pm$6.185& --0.090$\pm$0.018& 0.314$\pm$0.014& 36.609$\pm$9.312&   0.014$\pm$0.024& 0.092$\pm$0.028\\
$19<$V$<20$ &$V-I>1.6$& 28.649$\pm$5.302& --0.059$\pm$0.025& 0.302$\pm$0.025& 65.822$\pm$9.429&   0.029$\pm$0.012& 0.085$\pm$0.013\\
\noalign{\smallskip}
\hline\noalign{\smallskip}
\multicolumn{1}{l}{Galactic $b$} & &\multicolumn{3}{c}{Field} & \multicolumn{3}{c}{Cluster}\\
\noalign{\smallskip}
\hline\noalign{\smallskip}
V$<19$      &$V-I<1.6$& 55.828$\pm$5.611& --0.048$\pm$0.017& 0.202$\pm$0.011&	       &	  &	   \\
V$<19$      &$V-I>1.6$& 23.495$\pm$3.135& --0.063$\pm$0.021& 0.254$\pm$0.017& 98.707$\pm$ 0.  & --0.019$\pm$0.008& 0.089$\pm$0.   \\
$19<$V$<20$ &$V-I<1.6$& 56.149$\pm$3.678& --0.051$\pm$0.015& 0.291$\pm$0.010& 36.609$\pm$ 0.  &   0.007$\pm$0.025& 0.092$\pm$0.   \\
$19<$V$<20$ &$V-I>1.6$& 37.056$\pm$3.878& --0.066$\pm$0.018& 0.241$\pm$0.013& 65.822$\pm$ 0.  &   0.001$\pm$0.013& 0.085$\pm$0.   \\ 
\noalign{\smallskip}
\hline
\end{tabular}
\label{hist_redblue.tab}
\end{table*}

For the blue stars ($V_{\rm 555}-I_{\rm 814}<1.6$) in the brightest
sample, i.e.  $V_{\rm 555}<19$, it proved impossible to fit two
Gaussians both for $\mu_{\rm l}$ and $\mu_{\rm b}$. Since there are few stars in
these two  histograms we have carefully checked that the chosen
binning did not  effect the final result. The results are given in
Table \ref{hist_redblue.tab}.

The red samples, on the other hand, show a very strong central peak 
and broad wings which means that two Gaussians are needed to 
achieve a good fit. As before we first fitted the $\mu_{\rm l}$
distribution and then fixed the $\sigma$ and amplitude for the 
narrow Gaussian when fitting the $\mu_{\rm b}$ distribution. It is 
interesting to find that indeed the narrow Gaussian has a $\sigma=0.08$
arcsec per century, exactly the same as found when the full colour-range
was investigated. 

We also investigate the distributions for the magnitude range 19 to 20,
Fig. \ref{hist_redblue_v1920.fig}. Here, again, we see a rather broad
dominating dispersion in the blue while the red is dominated by
the narrow distribution, even if not as prominently as in the case of
the  brightest stars. 

The main change between the fit parameters for the sample split by
colour and for the whole sample, is that in $l$ the center for the
blue sample is significantly different from the red sample, and both
the blue and red centers are different to that found when the whole
sample is fitted. We suggest that this is due to some of the bright
blue stars being from a third stellar population, namely the Galactic
disk, which has an additional velocity component in $l$.

We note that the stars identified with the Galactic disk have a mean
proper motion in $l$ that is negative relative to that of the Galactic
bulge.  Disk dwarf stars at the magnitudes observed here should be
within a kpc or less from us (see e.g. Sadler et al. 1996).  If the
rotation curve of the disk (which is not well sampled for these type
of objects) differs somewhat from pure differential rotation then this
can explain our measured proper motions for stars at these distances.

\section{The proper motion of NGC 6528 and the velocity 
dispersion in the Bulge}
\label{sect:interpret}

The measured proper motions can be used to find the mean proper motion
of NGC 6528 relative to the Galactic bulge, and the bulge and cluster
velocity dispersions.

As discussed above, it appears that the bright blue stars contain a
Galactic disk component. Therefore we use the bright (V$<$19), red
(V-I$>$1.6) sample to get the final values for cluster and bulge
velocity dispersion and proper motion given in
Table \ref{tab:fin}. This then gives us a cluster proper motion
relative to the bulge of $<\mu_{\rm l}>$ = 0.006 and $<\mu_{\rm b}>$ = 0.044
arcsec/century.

\begin{table*}
\caption{Final values for velocity dispersion and mean proper motion.
An error of 0. indicates that that particular parameter was kept fixed
during the fitting procedure. All values are in arcsec century$^{-1}$}
\label{tab:fin}
\begin{tabular}{lllll}
\hline\noalign{\smallskip}
        & $\sigma_{\rm l}$        & $\sigma_{\rm b}$        & $<\mu_{\rm l}>$ & $<\mu_{\rm b}>$ \\
\noalign{\smallskip}
\hline\noalign{\smallskip}
Bulge   & 0.327$\pm$0.027 & 0.254$\pm$0.017 & --0.030$\pm$0.026 & --0.063$\pm$0.021  \\  
Cluster & 0.089$\pm$0.007 & 0.089$\pm$0.    & --0.024$\pm$0.007 & --0.019$\pm$0.008\\
\noalign{\smallskip}
\hline
\end{tabular}
\end{table*}

The velocity dispersion for the cluster, 0.08 arcsec per century,
translates to 24-30 km s$^{-1}$ for the upper and lower distance
limits to NGC6528 of 6.5 kpc and 8 kpc (Richtler et al. 1998).  (The
distance found by Ortolani et al. (1992) is 7.5 kpc.)

Globular clusters in the Galaxy have measured velocity dispersions
that range from a few km s$^{-1}$ to $\sim 20$ km s$^{-1}$, see Pryor
\& Meylan (1993) and Dubath et al. (1997). In M 31 at least two
globular clusters have measured velocity dispersions $>20$ km
s$^{-1}$, Dubath \& Grillmar (1997). Zoccali et al. (2001) found
$\sigma$=28 km s$^{-1}$ for NGC6553.  This result is very similar to
ours. Since most globular clusters in the Galaxy have significantly
lower velocity dispersion they concluded that their measured
$\sigma_{cluster}$ was dominated by measurement error. This is most
likely also the case for NGC6528.

Assuming the cluster velocity dispersion in NGC 6528 is dominated by
errors, we deconvolve this from the measured velocity dispersion for
the bulge to find the true bulge velocity dispersion.  Using the data
for the red sample (Table \ref{hist_redblue.tab} with $V_{\rm 555}<
19$) we get $\sigma_{\rm l~bulge}= 0.33\pm 0.03$ and $\sigma_{\rm b~bulge}=
0.25\pm0.02$ arcsec per century. These numbers are in good agreement
with the results for bulge giants found  by Spaenhauer et al. (1992),
$\sigma_{\rm l~bulge}= 0.32\pm 0.01$ and  $\sigma_{\rm b~bulge}= 0.28\pm0.01$
arcsec per century for their full sample of 429 stars.

These numbers give a  $\sigma_{\rm l}/\sigma_{\rm b}=1.32\pm0.16$, which is
identical, within the error estimates, to the 1.33 predicted for the
coordinates of NGC 6528 by the model of kinematics in the Galactic
bulge in Zhao (1996).

In their study of NGC 6553 Zoccali et al. (2001) derived
$\sigma_{\rm l~bulge}= 0.26\pm 0.03$ and $\sigma_{\rm b~bulge}= 0.21\pm0.02$
arcsec per century giving $\sigma_{\rm l}/\sigma_{\rm b}=1.24\pm0.17$. These
values are lower than found here, however, NGC 6553 is situated
further out from the Galactic centre than NGC 6528 and we should thus
expect $\sigma_{\rm l~bulge}$ to be a factor  $\sim 0.86$ lower than for
the coordinates of NGC 6528, see Zhao (1996) Table 6.
$\sigma_{\rm b~bulge}$ should remain roughly the same. Specifically the
model of Zhao (1996) predicts a $\sigma_{\rm l}/\sigma_{\rm b}=1.32$ at
$l=1,~b=-4$ and $\sigma_{\rm l}/\sigma_{\rm b}=1.09$  at $l=5,~b=-3$, which is
consistent, within the errors, to the values found here and in Zoccali
et al. (2001) for the bulge stars observed in the fields of NGC 6528
and NGC 6553 (which are situated close to the coordinates for which
Zhao's model makes its predictions).

We may thus conclude that these two new studies of the proper
motions of  Galactic bulge  stars confirm the predictions by models of
the kinematics in the  Galactic bulge. To our knowledge the current
work and that of Zoccali et al. (2001) are the first studies to
address the velocity dispersion, measured by proper motions, of bulge
stars below the horizontal branch.

We can use the measured proper motions ($<\mu_{\rm l}>=\sim +0.006$ and 
$<\mu_{\rm b}>=\sim +0.044$ arcsec per century) of the cluster 
relative to the Galactic bulge 
 along with a radial velocity of 210
km s$^{-1}$ (Carretta et al. 2001), cluster distance of 7.5 kpc (Ortolani et
al. 1992), solar peculiar velocity relative to the local standard of
rest of (u$_{\sun}$,v$_{\sun}$,w$_{\sun}$)=(10,5.25,7.17) km s$^{-1}$ (Dehnen
\& Binney 1998) and the rotational velocity of the local standard of
rest of 239 km s$^{-1}$ (Arp 1986) to calculate the absolute space velocity
components of NGC6528. These are ($\Pi$,$\Theta$,$W$)=
(-220, 17.4, 16.1)
km s$^{-1}$ ($\Pi$ points radially outwards from the Galactic
centre towards the cluster, 
$\Theta$ is oriented in the direction of Galactic rotation,
and W points towards the north Galactic pole, and we have made the
simplifying assumption that the cluster is at 
$(l,b)=(0,0)$).

In an attempt to estimate the internal errors on the derived
velocities we varied the proper motions for the cluster according
to the errors derived when fitting the histograms and we find that
the $\Pi$ velocity is unaffected by the errors while $\Theta$
varies between $\sim 8$ and $\sim 27$ km s$^{-1}$ and $W$
between $\sim 8$ and $\sim 23$ km s$^{-1}$. These should represent 
maximal internal errors in our analysis and thus we can conclude
that the velocities derived are fairly robust.

\section{The colour-magnitude diagram and the proper motions}

Our new colour-magnitude diagram is presented in
Fig. \ref{cmdall.fig}.  It contains all the photometry from both long
and short exposures, as described in Sect. \ref{sect:stellarphot}.

 Clearly visible is the stubby red horizontal branch. The red giant
bump, roughly 0.5 mags below the horizontal branch is less easily
isolated but appears to be present. To the left a ``plume'' of field
stars and/or blue stragglers in the cluster stretches up from the
turn-off region. The turn-off region itself is identifiable but
appears confused.  The exact position of the sub-giant branch is
difficult to establish since it is contaminated with field stars from
both the disk and bulge.  The red-giant branch rises rather vertically
but then appear to turn-over heavily to the red.  The red-giant branch
itself is very wide. This could be an indication of differential
reddening and/or large contamination from bulge stars roughly at the
same distance modulus as NGC 6528 but with a spread in both age and
metallicity.

The question now is whether the broad red giant branch and the fuzzy
sub-giant branch are mainly the result of differential reddening, see
e.g Ortolani et  al. (1992), or are primarily caused by contamination
of the colour-magnitude diagram by bulge stars (Richtler et
al. 1998).

\subsection{Applying the proper motions to isolate the cluster}

The idea is now to use the proper motions to separate the cluster and
bulge stars.  There are at least two interesting points here. The
first is to know how well we can ``decontaminate'' the red giant
branch of NGC 6528 from bulge stars.  Secondly, we want to find out
how many of the stars in the ``blue'' plume above the turn-off
realistically belong to the bulge, to the cluster, or to the
foreground disk. Also here we would like to know how well we can clean
the turn-off region from contaminating stars. The quality of the
turn-off region is a major limitation in the case of NGC 6528 for
determining a reliable relative or absolute age.

We use our Gaussian fits to the different regions of the
colour-magnitude diagram to estimate the proper motion cuts which
maximize the number of cluster stars relative to the number of bulge
stars, whilst still allowing enough cluster stars to make a good
cluster colour-magnitude diagram. Using different proper motion cuts
in different regions of the colour-magnitude diagram will affect the
relative numbers of cluster stars in each region but this does not
matter for comparison of the observed colour-magnitude diagram with
other globular clusters and with isochrones, where all we use is the
position of the cluster stars in the colour-magnitude diagram and not
their number density.

\begin{figure}
\resizebox{10.5cm}{!}{\includegraphics{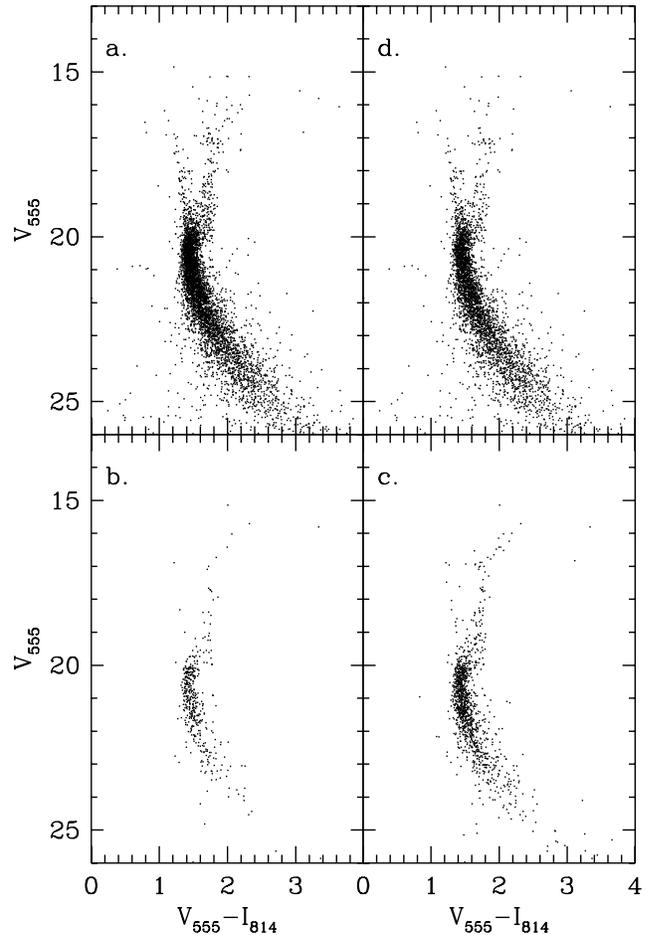}}
\caption[]{Illustration for WF2 of the effect on the colour-magnitude
diagram from different cuts in $\sqrt{\mu_{\rm l}^2+ \mu_{\rm b}^2}$.  All stars
are measured in both the new and old images and  satisfy the cuts
imposed in $\chi$ and sharpness variables (see
Sect. \ref{sect:stellarphot}).
Stars with fitting errors are excluded.  {\bf a} all stars, {\bf b}
stars with $\sqrt{\mu_{\rm l}^2+\mu_{\rm b}^2} > 0.18$, i.e. the rejected (mainly
bulge) stars, {\bf c} stars with $\sqrt{\mu_{\rm l}^2+\mu_{\rm b}^2} < 0.09$, and
{\bf d}  $\sqrt{\mu_{\rm l}^2+\mu_{\rm b}^2} < 0.18$ }
\label{cmdwf2.fig}
\end{figure}

In Fig. \ref{cmdwf2.fig}, we show the effect of various different
proper motion cuts imposed on the WF2 colour-magnitude diagram. 
We have previously found that the cluster has a $\sigma = 0.09$ 
arcsec per century. In the following we will use this $\sigma$
when defining the cuts to clean the colour-magnitude diagram.
In {\bf a} we show the full colour-magnitude diagram for WF2.
Plot {\bf c} and {\bf d} then shows the resulting
colour-magnitude diagram when a cut of $\sqrt{\mu_{\rm l}^2+\mu_{\rm b}^2} < 0.09$
and $\sqrt{\mu_{\rm l}^2+\mu_{\rm b}^2} < 0.18$, have been applied
respectively. For the most conservative cut the turn-off 
region becomes clean, although the sub-giant branch remains
somewhat confused. The number of stars in the giant branches, however,
becomes almost too small for quantitative work. Moreover, as seen
in Fig. \ref{hist_redblue.fig}, for the brighter magnitudes 
the red part of the colour-magnitude diagram, i.e. the giant 
branches, is dominated by cluster stars and a more generous 
cut can be allowed when cleaning
the colour-magnitude diagram. This is shown in {\bf c}. Here 
though the turn-off region again becomes too confused for
good work. 

\begin{figure}
\resizebox{10.5cm}{!}{\includegraphics{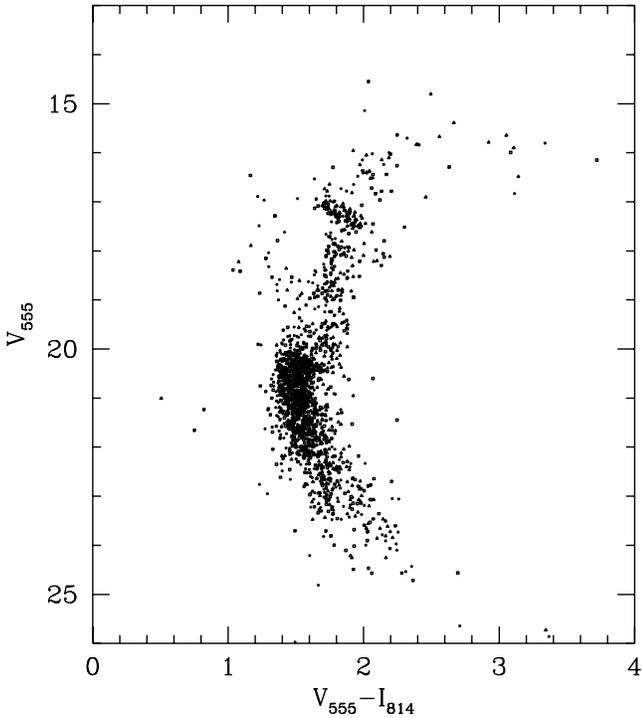}}
\caption[]{Colour-magnitude diagram from WF234 for the stars with 
$\sqrt{\mu_{\rm l}^2+\mu_{\rm b}^2} < 0.09$ for $V_{\rm 555}\geq 19$ and 
$\sqrt{\mu_{\rm l}^2+\mu_{\rm b}^2} < 0.18$ for the brighter stars }
\label{cmdsel.fig}
\end{figure}

Our cleaned colour-magnitude diagram, Fig. \ref{cmdsel.fig}, is
finally obtained by imposing the following cuts;
$\sqrt{\mu_{\rm l}^2+\mu_{\rm b}^2}<0.18 $,  for star with $V_{\rm
555}<19$ and $\sqrt{\mu_{\rm l}^2+\mu_{\rm b}^2}<0.09 $ for the fainter stars.

\begin{figure}
\resizebox{10.5cm}{!}{\includegraphics{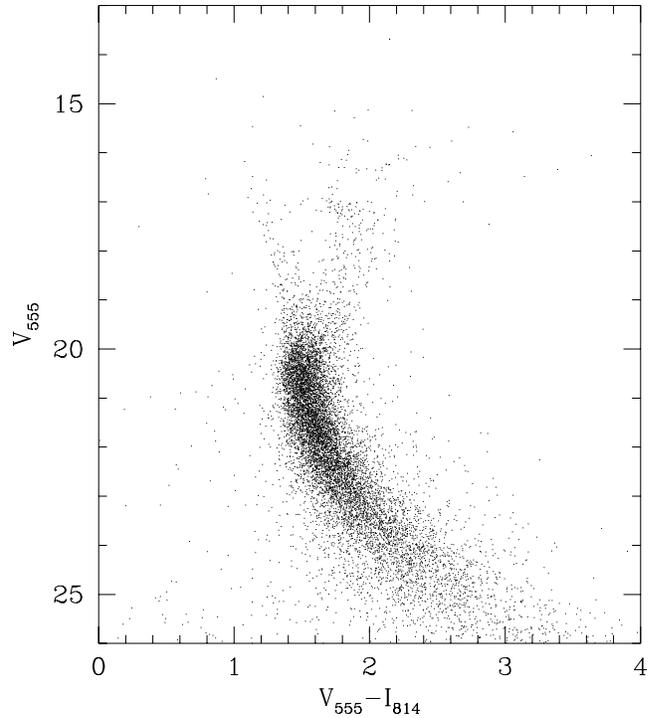}}
\caption[]{Colour-magnitude diagram from WF234 for stars rejected
as mainly bulge stars due to their proper motions are included,
i.e. $\sqrt{\mu_{\rm l}^2+\mu_{\rm b}^2} > 0.18$ }
\label{cmdrej.fig}
\end{figure}

In Fig. \ref{cmdwf2.fig} {\bf b}, finally,
we show the stars that have $\sqrt{\mu_{\rm l}^2+\mu_{\rm b}^2} > 0.18$.
These are mainly bulge stars. Compare also the various histograms. 

Figure \ref{cmdrej.fig} shows the colour-magnitude diagram for the
stars that have $\sqrt{\mu_{\rm l}^2+\mu_{\rm b}^2} > 0.18$, i.e.  mainly bulge
stars. This diagram is now based on all three WFs.  This
colour-magnitude diagram has large spreads everywhere. Particularly
noteworthy is the plume of stars that emanates from the turn-off
region as well as the extremely fuzzy appearance of the regions around
the horizontal branches, indeed almost a lack of horizontal branch.
This colour-magnitude diagram should be compared to that of Baade's
window, see e.g. Feltzing \& Gilmore (2000) and Figure 2 in Holtzman
et al. (1998). We note also that the red giant stars show a branch
that is turning over significantly.

\begin{figure*}
\resizebox{16cm}{!}{\includegraphics{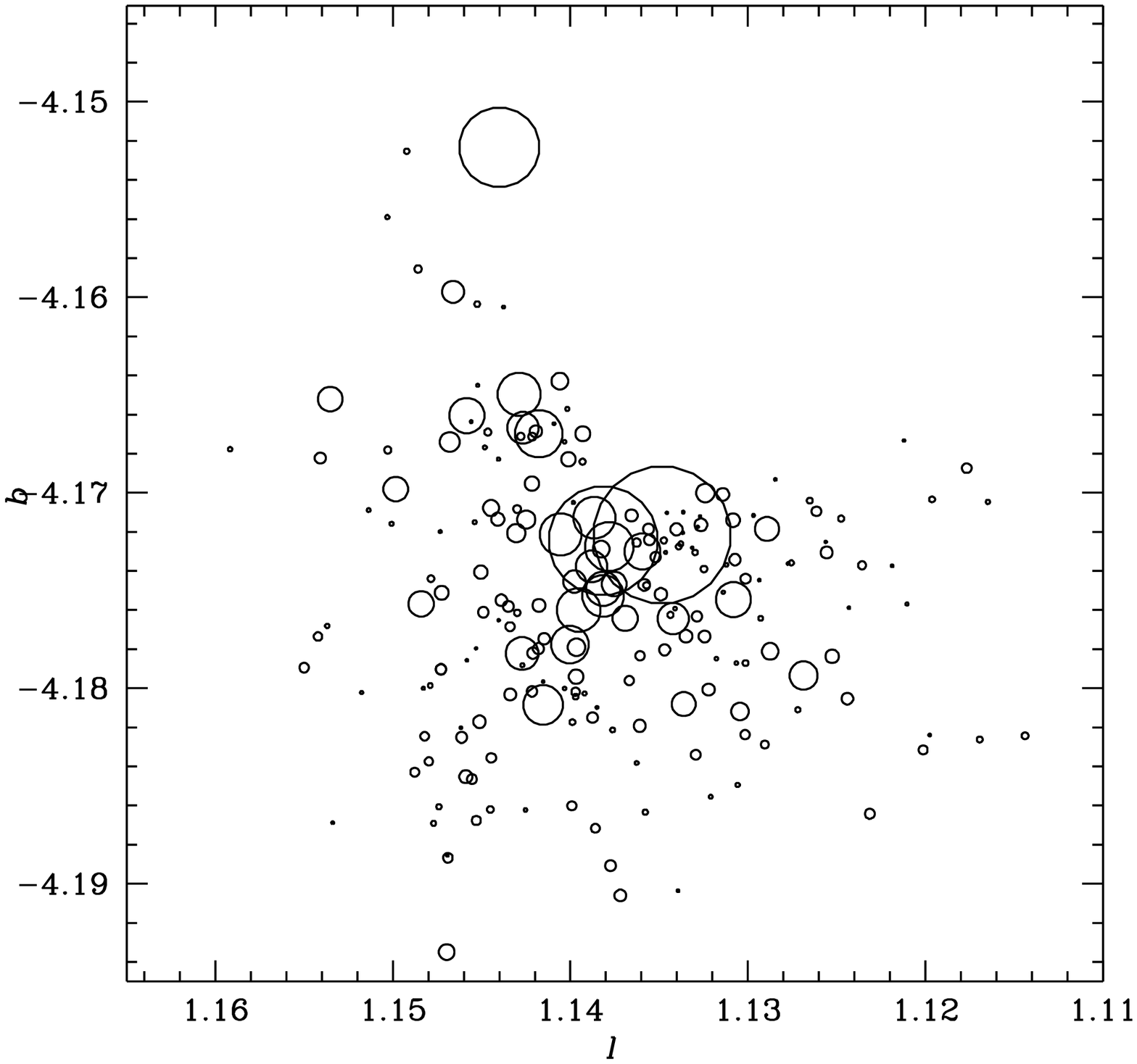}
\includegraphics{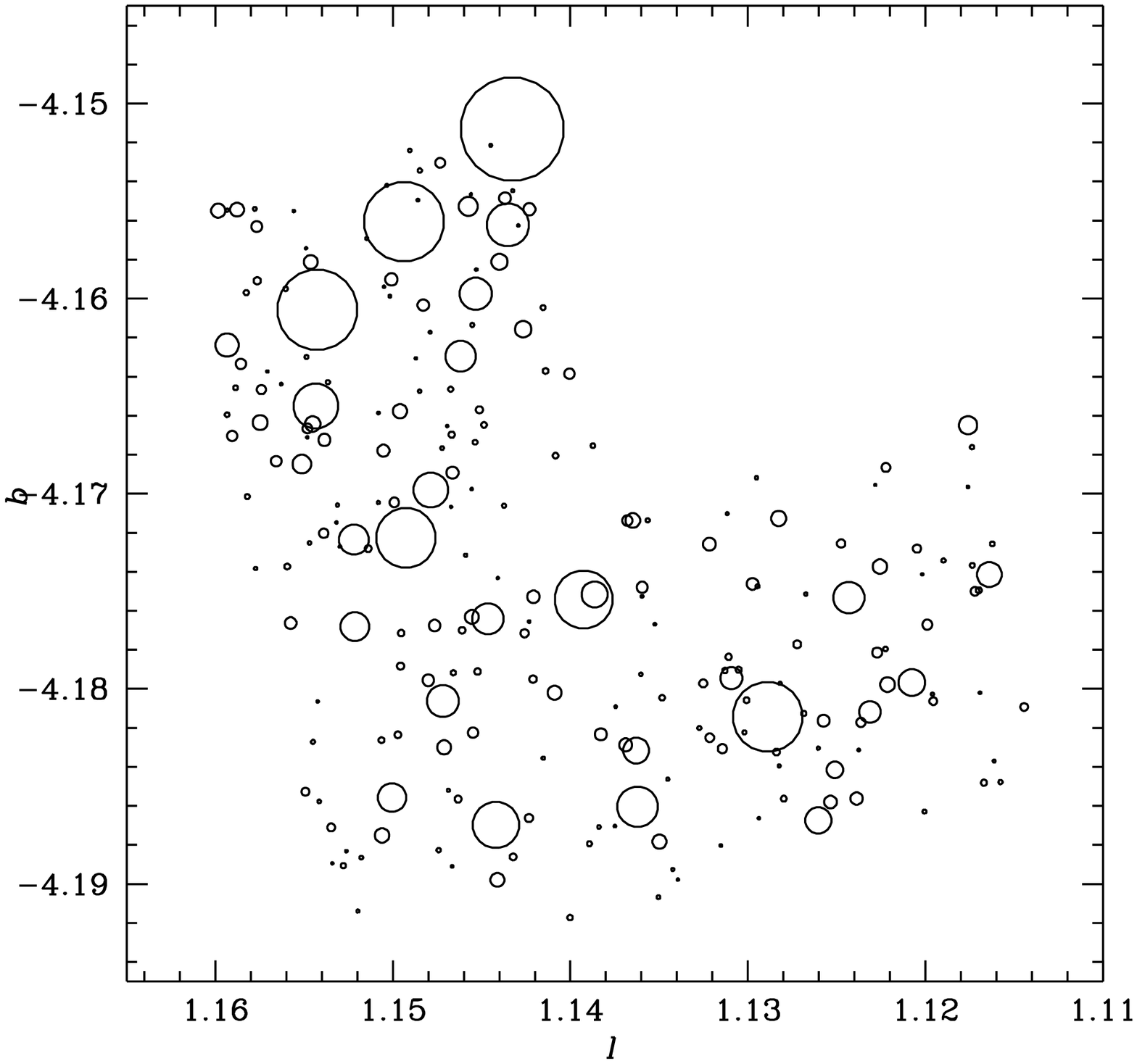}}
\caption[]{
Positions on the sky for {\bf a} stars with
 $V_{\rm 555}<19$, $\sqrt{\mu_{\rm l}^2+\mu_{\rm b}^2} < 0.09$, and
$V_{\rm 555}-I_{\rm 814}>1.6$. This should be predominantly 
cluster stars. In
{\bf b} stars with $V_{\rm 555}<19$, $\sqrt{\mu_{\rm l}^2+\mu_{\rm b}^2}\geq 0.18$
and for all colours. This selection should primarily give us 
bulge and foreground disk stars. The sizes of the symbols code 
the magnitude of the stars
}
\label{pos.fig}
\end{figure*}

For stars brighter then $V_{\rm 555}=19$ Fig. \ref{pos.fig} shows how
the stars, included and excluded, from the colour-magnitude diagram
are distributed on the sky. The first panel shows the stars that have
the highest probability to belong to the globular cluster, i.e. small
proper motions and redwards of 1.6 in colour.  The second shows the
stars that are most likely to belong to either the bulge or to be
foreground disks stars. These plots give further  support for our
definition of cuts in proper motion when defining the stars that
belong to the cluster.  Figure \ref{pos.fig} {\bf a} show a fairly
concentrated structure, which tapers of at a certain  radius. Note
that the detection of stars in the very centre is limited because here
we are only using the short exposures since the long were too crowded
for good positions. Figure \ref{pos.fig} {\bf b} on the other hand
shows a much more even distribution of stars.

\subsection{Differential reddening towards NGC 6528}

\begin{figure*}
\includegraphics[angle=-90,width=12cm]{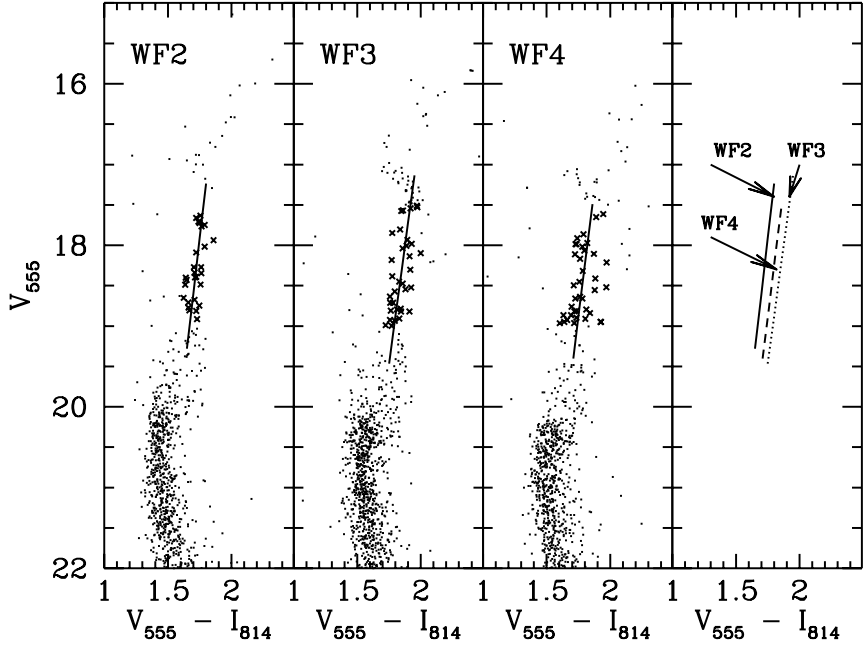}
\caption[]{Colour-magnitude diagrams for WF234 as indicated. Stars
used to provide a linear-least square fit to the red giant branch are shown as 
$\times$.
The fits are indicated by full lines for the individual chips and in the last
panel with full (WF2), dotted (WF3) and, dashed (WF4) lines. }
\label{redall.fig}
\end{figure*}

We quantify how much of the apparent spread in the colour-magnitude
diagram in Fig. \ref{cmdsel.fig} is due to differential reddening by
fitting the ``straightest'' portion of the red giant branch for each
chip using a linear least squares fit. This is shown in Fig. 
\ref{redall.fig}.  In the final
panel the fits for the three different WFs are compared. From this it
is clear that, in the {\sl mean}, the reddening differs between the
three chips such that WF2 has the smallest reddening and WF3 has the
largest. These reddening estimates are obtained only for stars 
that most likely belong to the cluster, i.e. the same cuts are
imposed in all the following plots as we did in Fig. \ref{cmdsel.fig}.

We also consider below whether differential reddening is significant within
each chip.

\paragraph{WF2 }The
giant branch in the colour-magnitude diagram derived from WF2 is quite
tight and well defined. We take this as an indication that the
differential reddening over this chip is small and that no correction
for differential reddening within the chip is needed.

\begin{figure*}
\includegraphics[angle=-90,width=12cm]{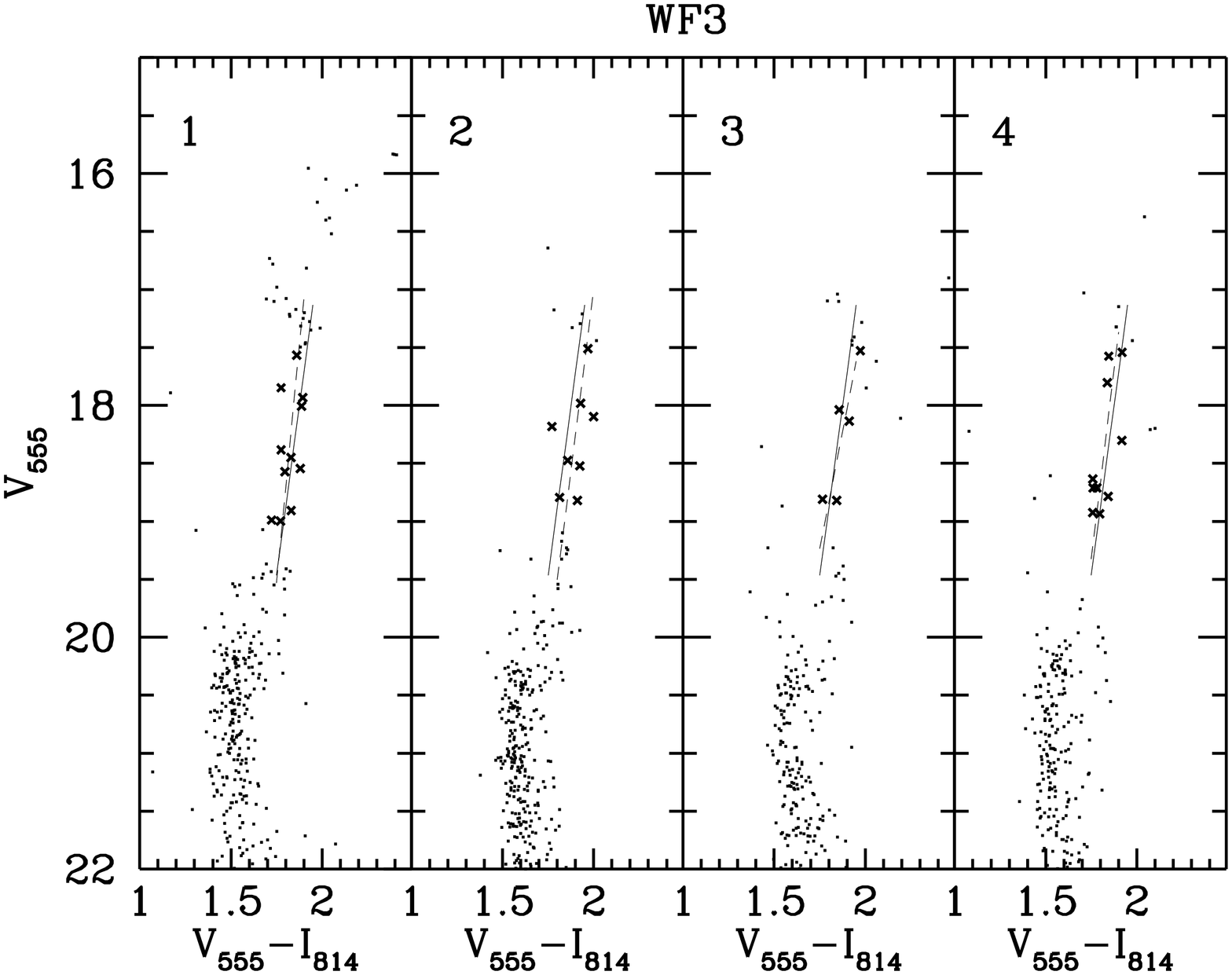}
\includegraphics[width=12cm]{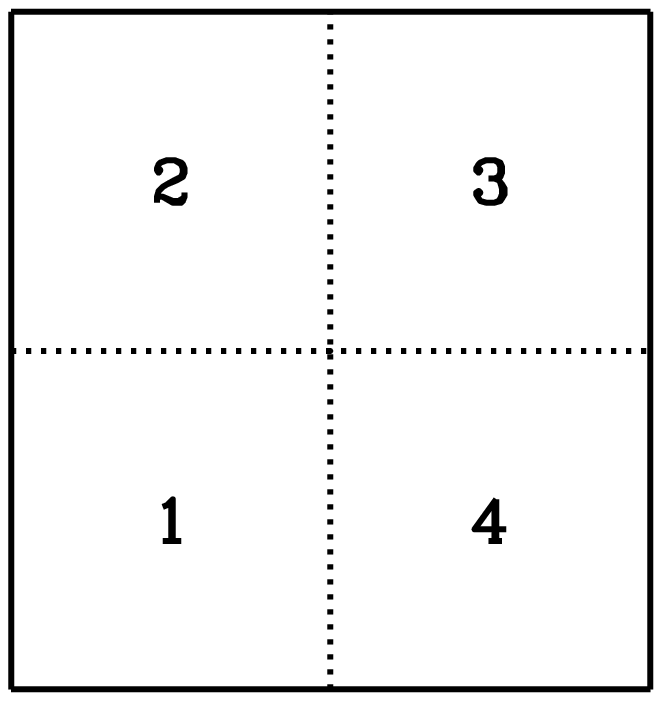}
\caption[]{Colour-magnitude diagrams for the four quadrants of 
WF3. The division of the chip and labeling of the resulting
colour-magnitude diagrams are shown to the right }
\label{redwf3.fig}
\end{figure*}

\begin{figure}
\resizebox{\hsize}{!}{\includegraphics{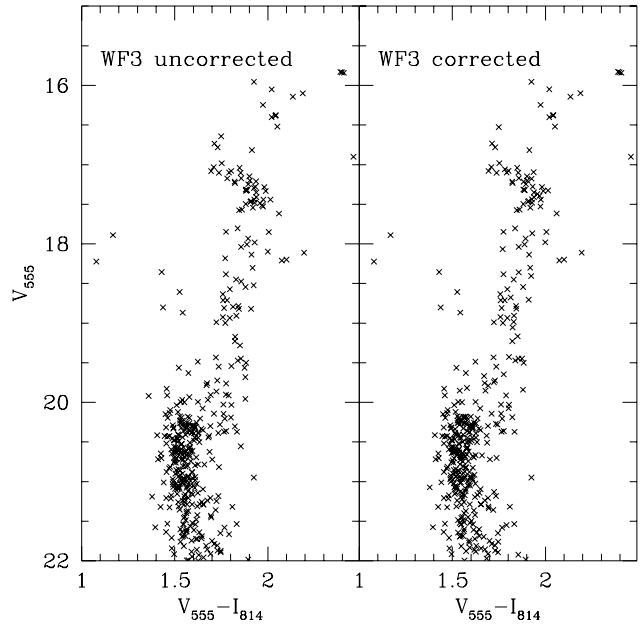}}
\caption[]{The effect achieved by correcting the second quadrant of WF3
for differential reddening with respect to the other three quadrants
(see Fig. \ref{redwf3.fig}).
A reddening of $E(V_{\rm 555}-I_{\rm 814})=0.0447$ was applied}
\label{cmdwf3_corr.fig}
\end{figure}

\paragraph{WF3} We divide the image into four sections and 
construct the colour-magnitude diagrams for each of them in Fig. 
\ref{redwf3.fig}. 
Again stars on the cleanest portion of the giant branch are selected
and fitted with a straight line. First we fit it to all the stars on
the chip (full line). Then for each quadrant of the image (dashed
lines). From this it is clear, although the number statistics are low,
that quadrant 1, 3, and 4 all are close to the mean value, while
quadrant 2 has a larger reddening than the rest. As the scatter around
the fitted line for that part of the image is small we do not further
divide the image into smaller sections but will correct stars in
quadrant 2 according to this shift, see Table
\ref{red.tab}. First we found the shift in $V_{\rm 555} - I_{\rm 814}$.
Then we used Holtzman et al. (1995b) Table 12 to find the corresponding
$\Delta V_{\rm 555}$ that should be applied to remove the colour
difference.
Figure \ref{cmdwf3_corr.fig} shows the 
WF3 colour-magnitude diagram corrected for
differential reddening.

\begin{table}
\caption[]{Differential reddening corrections for the quadrants
of WF3 to the mean reddening of WF2. In each
case the values indicate how much the stars were moved in order for their
fiducial line for the red giant branch to coincide with that of the
bluest one within the WF. We used the extinction laws and values
in Holtzman et al. (1995b)}
\begin{tabular}{llllll}
\hline\noalign{\smallskip}
     Quadrant & $\Delta({\rm V}_{\rm 555}-{\rm I}_{\rm 814})$ & 
$\Delta {\rm V}_{555}$ & \\
\noalign{\smallskip}
\hline\noalign{\smallskip}
 1,3,4  &  --0.106 & --0.273\\
 2      & --0.151 & --0.388 \\
\noalign{\smallskip}
\hline
\end{tabular}
\label{red.tab}
\end{table}

\begin{figure*}
\includegraphics[angle=-90,width=12cm]{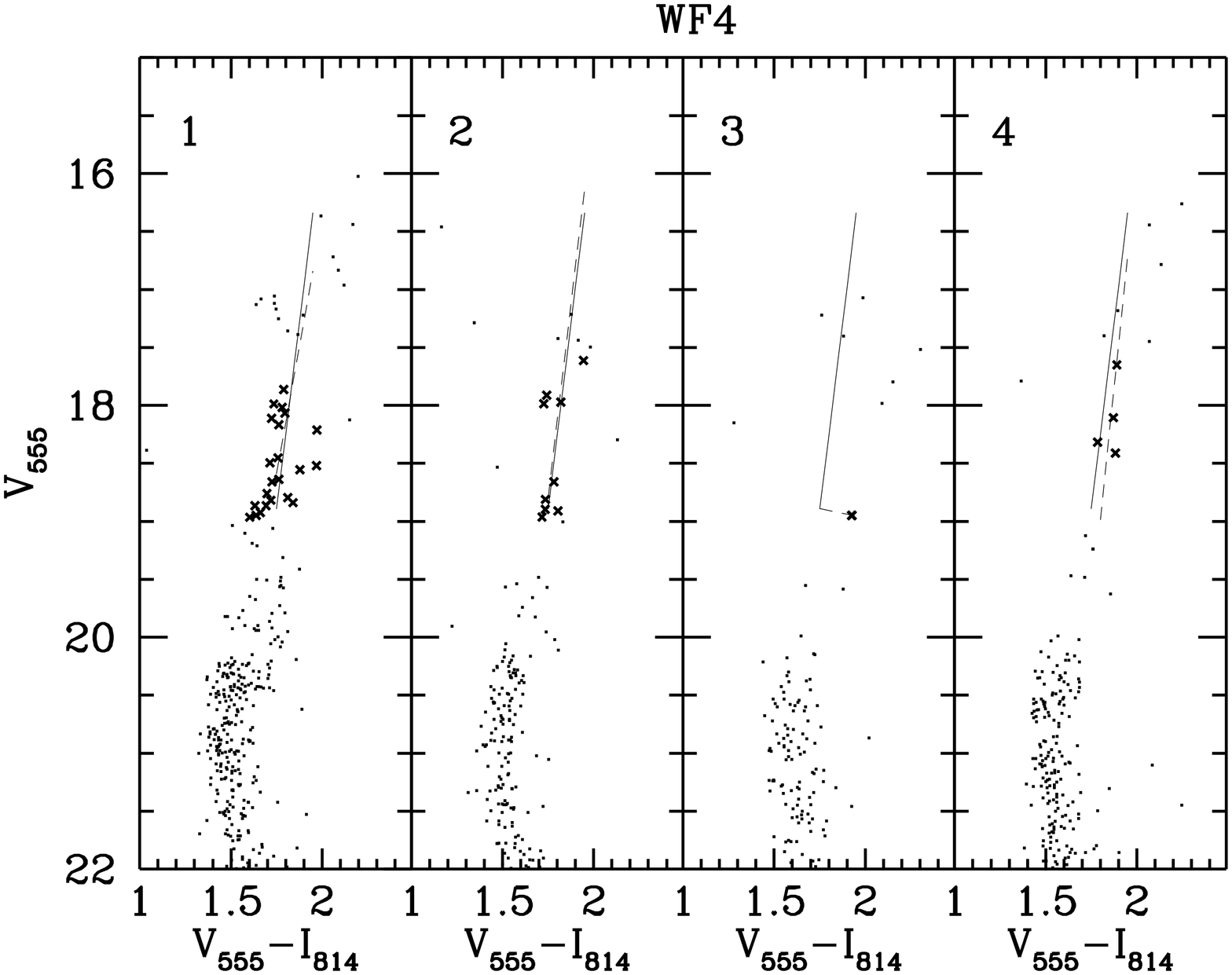}
\includegraphics[width=12cm]{h3031f14.ps}
\caption[]{Colour-magnitude diagrams for the four quadrant of 
the WF4 image. The division of the chip and labeling of the resulting
colour-magnitude diagrams are shown to the right }
\label{redwf4.fig}
\end{figure*}

\begin{figure*}
\includegraphics[angle=-90,width=12cm]{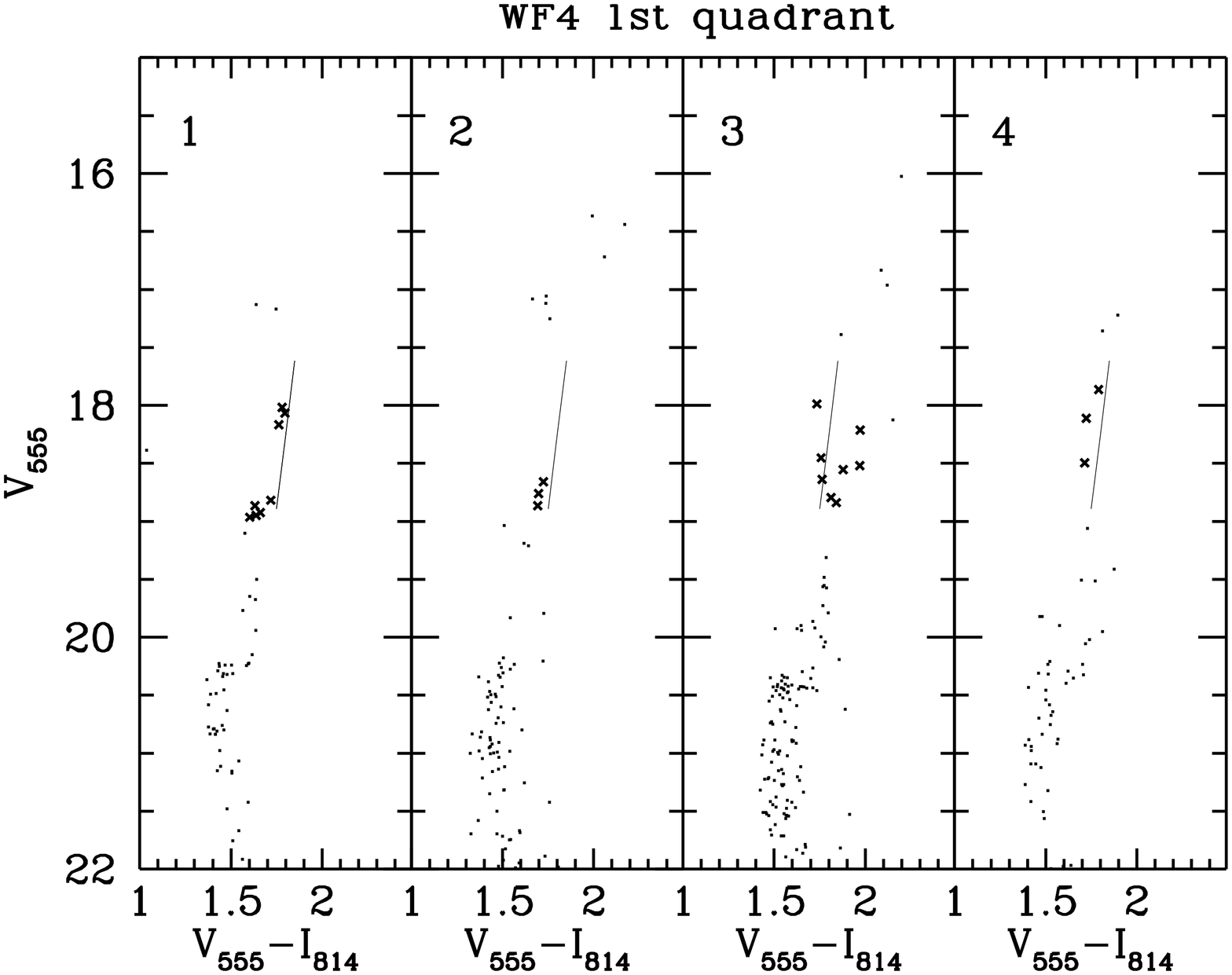}
\includegraphics[width=12cm]{h3031f14.ps}
\caption[]{Colour-magnitude diagrams for the four sub-quadrants of
the first quadrant of WF4 image.  The division of the chip and
labeling of the resulting colour-magnitude diagrams are shown to the
right }
\label{redwf42.fig}
\end{figure*}

\paragraph{WF4} The colour-magnitude diagram for this chip has the largest 
spread on the giant branch. Figure \ref{redwf4.fig} shows the result
for the four subsections. 
For quadrant 3 on the chip nothing can be
said since the number statistics is too low.  Quadrant 2 appears to be
well lined up with the mean value for the chip and quadrant 4 has
somewhat larger reddening than quadrant 2.  The colour-magnitude
diagram for the first quadrant, however, has a remaining 
large scatter and we
further subdivide this quadrant into four 
sub-quadrants, Fig. \ref{redwf42.fig}. 

This shows that the largest scatter emanates from sub-quadrant 3 and
that the three remaining sub-quadrants have a reddening that is less
than the mean reddening for the first quadrant. However, because of
the complexity found for the differential reddening we will omit WF4
from further discussions.

\begin{figure}
\resizebox{10.5cm}{!}{\includegraphics{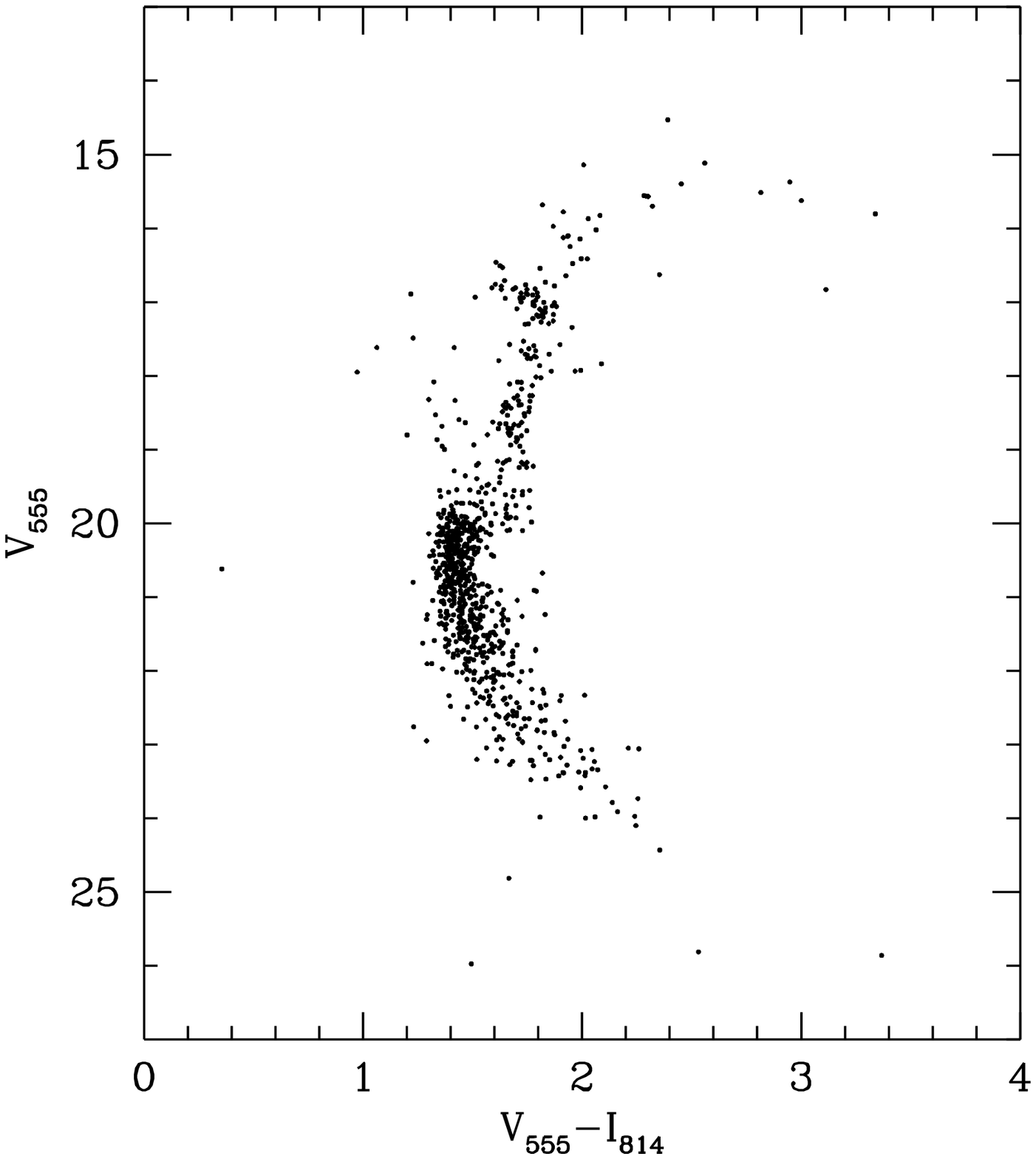}}
\caption[]{Final colour-magnitude diagram based on the data from WF2 and 
WF3 corrected for differential reddening. Cuts in   $\sqrt{\mu_{\rm l}^2+\mu_{\rm b}^2}$ 
as in Fig. \ref{cmdsel.fig}}
\label{cmd_final.fig}
\end{figure}

Our full final colour-magnitude diagram, using data from WF2 and WF3
corrected for differential reddening, is shown in Fig. \ref{cmd_final.fig}.

We observe that the red giant branch is more pronounced when the reddening
corrections have
been applied (see also  Richtler et al., 1998). The red 
subgiant branch bump, even though it is not a strong feature in
our colour-magnitude diagram, appear more rounded and well-defined.
Heitch \& Richter (1999) used the lumpiness of
this feature to assess the quality of their differential
dereddening. Thus we take the improvement in our corrected colour-magnitude
diagram of this feature as an indication that the  differential reddening that
has been applied is the correct one. 

\subsection{The tilt of the HB and the curved asymptotic giant branch}

For a long time it has been known that one of the characteristics of
metal-rich globular clusters in comparison with the metal-poor
clusters is the presence of a strongly curved red asymptotic giant
branch in the optical. This effect is caused by the extra line
blanketing provided by the numerous molecular lines, e.g. TiO, present
in the spectra of cool metal-rich giant stars.  The complexity of such
spectra is illustrated by e.g. Fig. 2 in Ortolani et al. (1991).  In
Ortolani et al. (1992) the optical colour-magnitude diagram for NGC
6528 shows just this curved structure.  Richtler et al. (1998) were
able to define a sample of NGC 6528 stars extending all the way out to
$V-I \sim 6.5$, showing an exceptionally curved red giant branch and
asymptotic giant branch (AGB) which after $V-I \sim 4$ progressively
deviates from the predictions from stellar evolutionary tracks (see
their Fig. 7).

Fig. \ref{cmd_final.fig} shows the full extent of our AGB.  There are
a few red stars that are fainter than the majority of the AGB. These 
stars could be members of the bulge, but
could also, obviously belong to the cluster and be in a region
which has a larger differential reddening which is on such a small
scale that our previous investigation could not detect it.

An upper envelope for the AGB has been found, but since we here are
only using material from two WFPC2 chips the field of view is small we
are content with saying that our results agree well with those of
Richtler et al. (1998) for the upper envelope, see their Fig. 7. They
also find a number of stars with $V-I> \sim 4$. We find no such stars
in our proper motions selected sample. The reason for this could
either be that the stars are saturated in our I-band images or
that they happen to
be outside our field of view. We are not
in a position to be able to further distinguish between these
possibilities.

\section{The age of NGC 6528}

Armed with the cleanest colour-magnitude diagram so far for the globular
cluster NGC 6528 we are in a position to address its age
using both  stellar isochrones as well as in relation to other well studied
metal-rich globular clusters.

\begin{figure}
\resizebox{10.5cm}{!}{\includegraphics{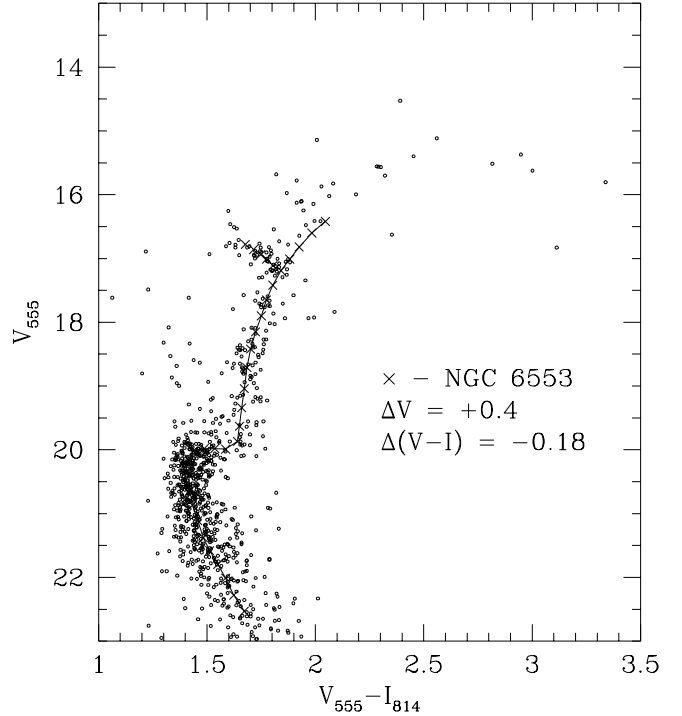}}
\caption[]{Comparison of our final colour-magnitude diagram 
with the ridge line derived for NGC 6553 by Zoccali et al. (2001). 
Their ridge line is showed as a set of connected $\times$ symbols. 
Both the mains-sequence,
red giant branch as well as the horizontal branch coincide well. The ridge 
line of NGC 6553 was moved $+0.4$ mag in $V_{\rm 555}$ and $-0.18$ in
$V_{\rm 555}-I_{\rm 814}$ }
\label{cmd_w_zoc.fig}
\end{figure}

\begin{figure}
\resizebox{10.5cm}{!}{\includegraphics{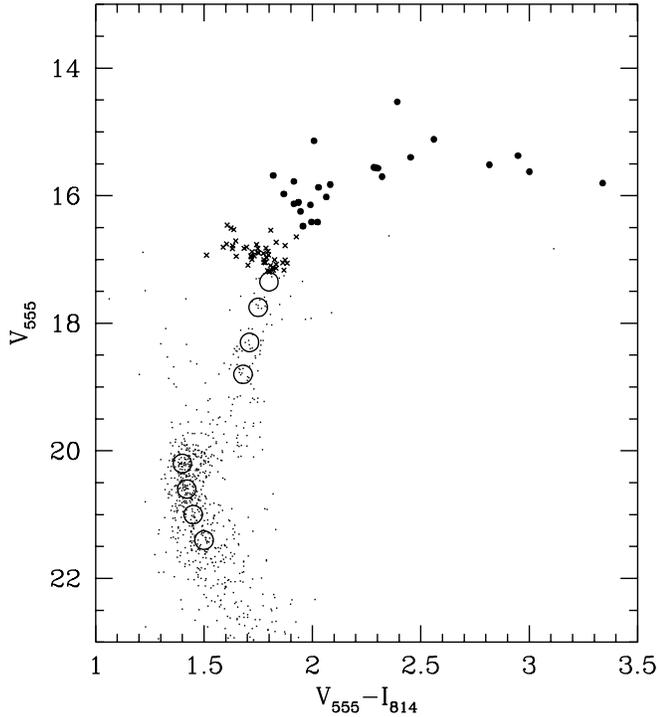}}
\caption[]{Colour-magnitude diagram with fiducial points indicated by
large open circles. The horizontal branch stars are indicated by
$\times$ and the AGB stars with $\bullet$. Note that
no points are defined for the lower part of the red giant branch, see text
for a discussion of this. Cuts in $\sqrt{\mu_{\rm l}^2+\mu_{\rm b}^2}$ as in
previous plots }
\label{cmd_w_ridge.fig}
\end{figure}

\subsection{The metallicity of NGC 6528}

Since age and metallicity are degenerate in colour-magnitude  diagrams
it is important to have a good estimate of  the metallicity if we want
to use stellar isochrones to find the absolute age of the cluster. In
Table \ref{ngc6528.tab}, we summarize the current  metallicity and
iron abundance  estimates  for NGC 6528 available in the literature.

Due to the faintness of metal-rich globular clusters detailed stellar
abundance studies of individual stars in these clusters have been
few. The results of the first such study was reported in Barbuy
(1999) who obtained  spectra of one, very cool ($T_{\rm eff} =3600 $
K) giant star in NGC 6528 and derived  a [Fe/H] of -0.6 dex, [Ca/Fe] =
0.0 dex and [Ti/Fe] = +0.6 dex.

Using the HIRES spectrograph on KECK  Carretta et al. (2001) and Cohen
et al. (1999), in two accompanying papers, derive iron abundances as
well as abundances for a large number of other elements for four stars
in NGC 6528 and five stars in NGC 6553.   Carretta et al. (2001) found
that all four stars in NGC 6528 show  very similar [Fe/H] ($+0.05,
+0.08, + 0.09, +0.04$ dex respectively) thus their final [Fe/H]
estimate for the  whole cluster appears very robust. The accompanying
study of NGC 6553 and the discussion of the  [Fe/H] for
NGC 6553 appear to indicate that a total error in [Fe/H] on
the order of 0.1 dex appear a reasonable estimate (see detailed
discussions  in Carretta et al. 2001).

See also their detailed discussion of the problems with analysis of as
cold giants. It appears that the the disagreement between the two
studies could be due to that different  temperature scales have been
used. Since the Carretta et al. (2001) study is the larger one and
also guided by their discussion on the temperature and errors from
other sources we here give higher weight to that  study for
determination of stellar abundances in NGC 6528. However, further
independent studies of the stellar abundances in this cluster should
be undertaken to solve this issue.

Further Carretta et al. (2001) found that the $\alpha$-elements  Si
 and Ca in NGC 6528 show large excesses compared to the solar values,
 while Ti and Mg appear to be solar. This type of abundance pattern is
 reminiscent of that observed for stars in the Galactic bulge
 (McWilliam \& Rich 1994) and might be indicative of a rapid chemical
 enrichment process prior to the formation of the stars observed. The
 exact interpretation of these data is, however, pending.

In Carretta et al. (2001) the cluster membership for the four stars
studied was ascertained by observing only horizontal branch stars.
The stellar spectra were also used to derive radial velocities for the
program stars and their velocities further confirmed the cluster
membership for the four targets.

Having assessed the currently available abundance information for NGC
6528 we find it safe to assume that the cluster is probably as
metal-rich as the sun and is enhanced, at least in some,
$\alpha-$elements.

\subsection{Comparison with NGC 6553}

In Fig. \ref{cmd_w_zoc.fig} we compare the ridge line for NGC 6553
found by Zoccali et al. (2001) with our colour-magnitude diagram.

To fit the colour-magnitude diagram for NGC 6528 with the ridge line
of NGC 6553 we need to move the ridge line from Zoccali et al. (2001)
by $0.4$ in $V_{\rm 555}$ and $-0.18$ in $V_{\rm 555} -I_{\rm
814}$. If the two clusters have the same metallicity (as indicated by
the recent detailed abundance analyses) this nice fit indicates that
their ages are very close too. We thus find that NGC 6553 has
been more reddened than NGC 6528, i.e. the negative shift applied in
$V_{\rm 555} -I_{\rm 814}$. Using Table 12 in Holtzman et al. (1995b)
we find that shift in $V_{\rm 555} -I_{\rm 814}$ corresponds to a
shift in $V_{\rm 555}$ of $-0.46$.  Thus NGC 6553 appears to be
marginally closer to us that NGC 6528.  However, it should be
remembered that both in Zoccali et al. (2001) and in this work the
colour-magnitude diagrams have been corrected for differential
reddening so the interpretation of such shifts is less clear in terms
of distance modulus.   The $\Delta E(B-V)$, as measured here,
between the two clusters is $\sim 0.15$, using Table 12 in Holtzman et
al. (1995b).

The agreement between the NGC 6553 ridge line and our data is very
good. At the brightest end the NGC 6553 ridge line appears to fall
slightly below the NGC 6528 data. However, since the NGC 6553 data
does not go as bright as the NGC 6528 data one should not draw any
conclusions regarding the relative metallicities of the clusters from
this.

This comparison does, as has also been reported in e.g. Ortolani et
al.  (1995), show that these two clusters have indeed very similar
ages.

\subsection{Age from fitting stellar isochrones}

Since NGC 6528 is found to be enhanced, at least  in some,
$\alpha-$elements we  compare our fiducial points with that of
theoretical stellar isochrones from Salasnich et al. (2000) in which
$\alpha$-enhancement has been included. To facilitate the  comparison
with the stellar isochrones we define a set of fiducial points which
are shown in Fig. \ref{cmd_w_ridge.fig} and tabulated in
Table \ref{ridge.tab}.

\begin{table}
\caption[]{Fiducial points  for NGC 6528 }
\begin{tabular}{lll}
\hline\noalign{\smallskip}
$V_{\rm 555}-I_{\rm 814}$ &$V_{\rm 555}$\\
\noalign{\smallskip}
\hline\noalign{\smallskip}
1.5   &  21.4   \\
1.45  &  21     \\
1.42  &  20.6   \\
1.4   &  20.2   \\
1.68  &  18.8   \\
1.71  &  18.3   \\
1.75  &  17.75  \\
1.8   &  17.35  \\
\noalign{\smallskip}
\hline
\end{tabular}
\label{ridge.tab}
\end{table}
The fiducial points are marked with large circles and trace the upper
main-sequence as well as the upper part of the red giant branch. We
have deliberately not defined any point for the lower red giant branch
or for the sub-giant branch as we feel that these regions are less
well defined and thus that any fiducial point might be misleading. We
have also marked those stars that we will explicitly show in the
diagrams where we fit isochrones. It is fairer to show all the AGB and
horizontal branch stars rather than try to represent them with a
fiducial line since then each reader is able to judge for themselves
the goodness of the fit.

\begin{figure}
\resizebox{\hsize}{!}{\includegraphics{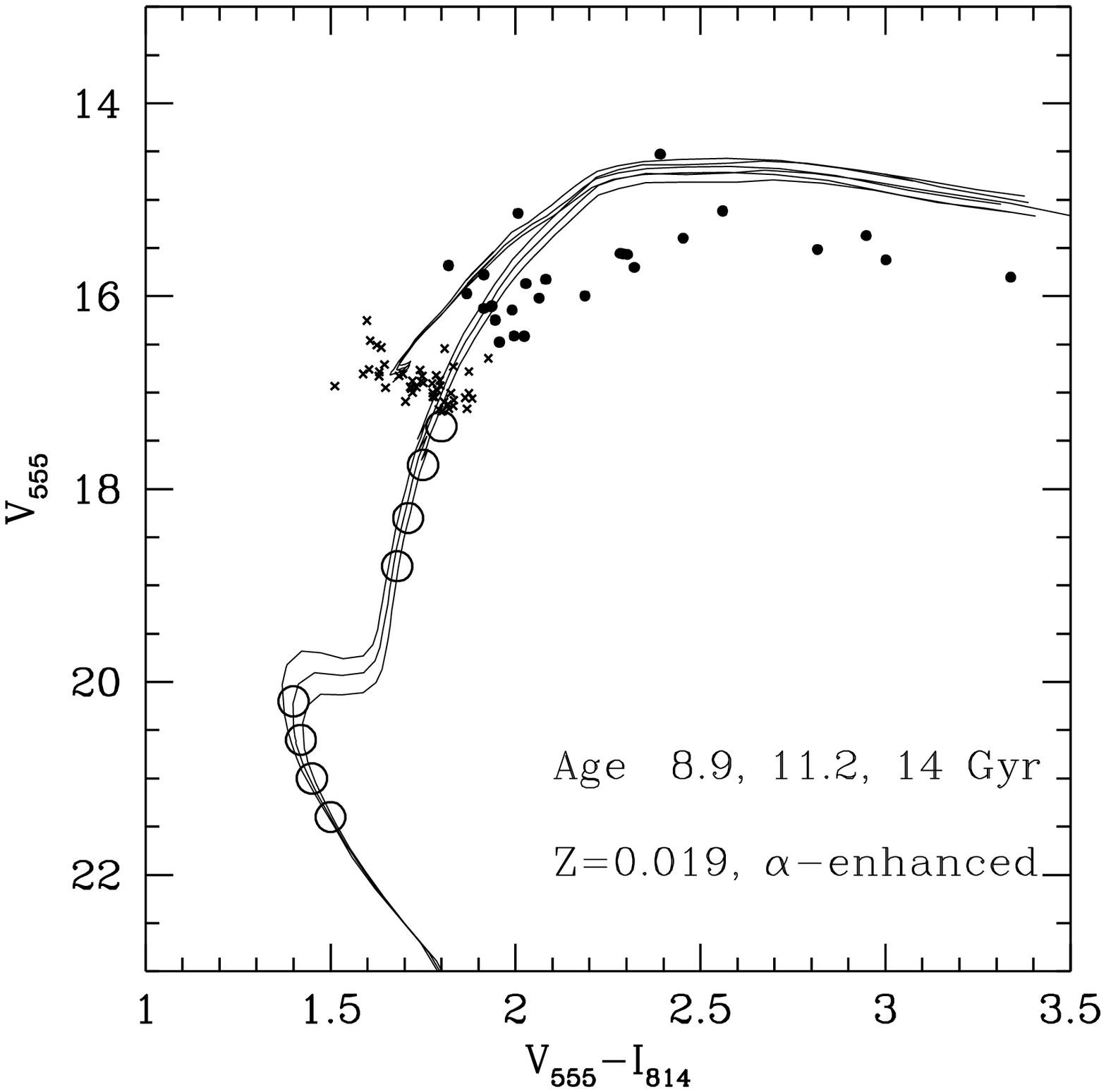}}
\resizebox{\hsize}{!}{\includegraphics{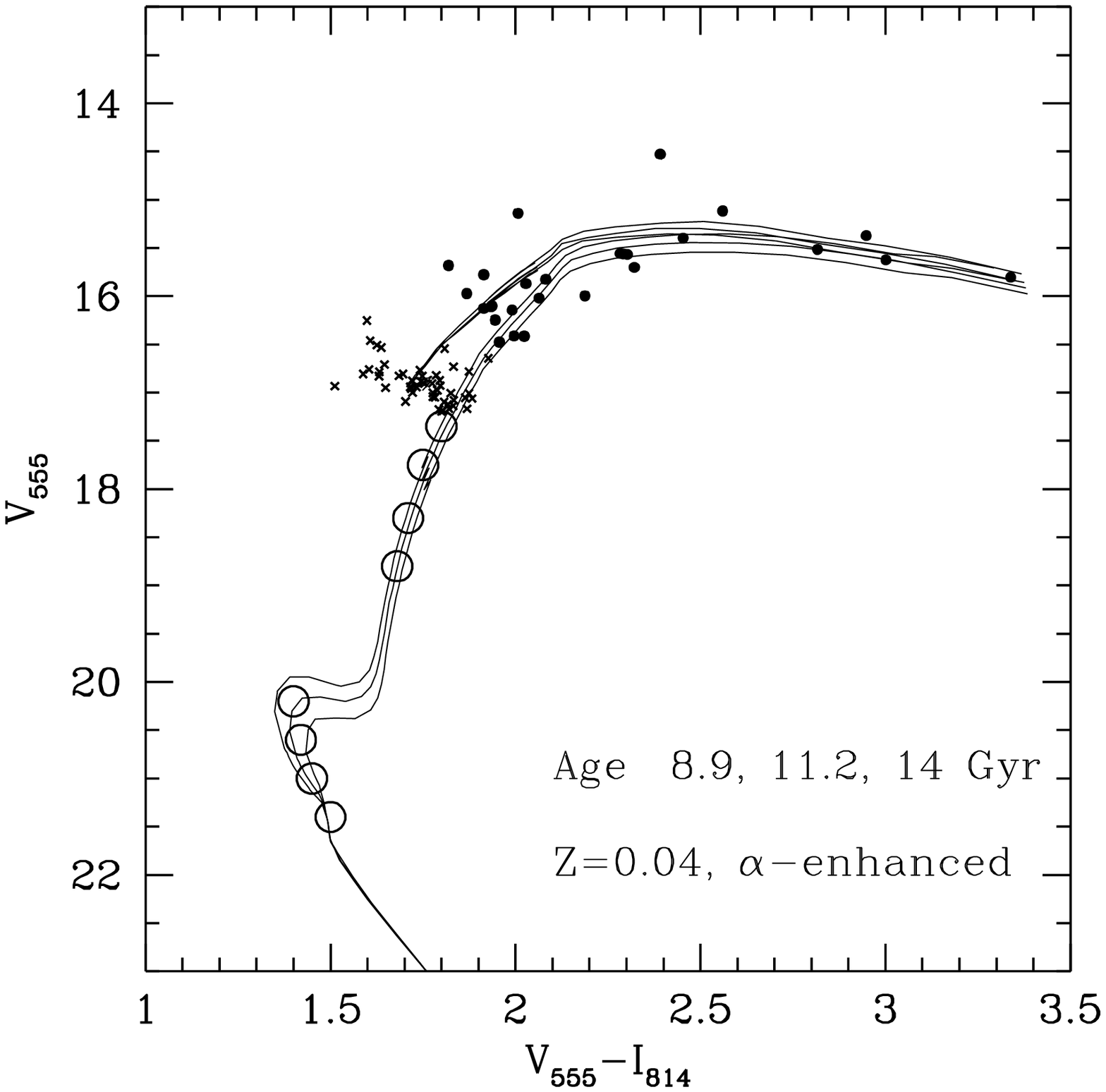}}
\caption[]{Fiducial points for NGC 6528 together with stellar
isochrones from Salasnich et al. (2000). Isochrones are for
$Z=0.019$ and $Z=0.040$. Horizontal
branch stars and AGB stars are coded as previously }
\label{isofit.fig}
\end{figure}

Examples of  fits are shown in Figs. \ref {isofit.fig}.

In the case of $Z=0.019$ we moved the isochrones by $\Delta(V_{\rm
555})=15.95$ and $\Delta(V_{\rm 555}-I_{\rm 814})=+0.63$. The turn-off
is well represented by the 11 Gyr isochrone and the horizontal branch
is well matched too. However, all the isochrones are  brighter than
the AGB. In order for $Z=0.019$ isochrones to fit our data on the AGB
we would need to increase the distance modulus and the best fitting
isochrone would then be very young, younger than 9 Gyr. The horizontal
branch would not be well fitted either. Thus it appears unlikely that
our data could be well fitted with $\alpha$-enhanced isochrones with
$Z=0.019$.

For the $Z=0.040$ isochrones we moved them with the following amount
$\Delta(V_{\rm 555})=15.95$ and  $\Delta(V_{\rm 555}-I_{\rm
814})=+0.655$. Here the AGB is much better reproduced and both
turn-off and horizontal branch can be well fitted simultaneously. The
11 Gyr isochrone appears to fit best. However as this fit cannot be
rigorous  due to the limitations in the data the estimated error bar
on this must be rather large, perhaps up to 2 Gyr.

These $\Delta(V_{\rm 555}$ and $\Delta(V_{\rm 555}-I_{\rm 814})$
correspond to (using Table 12 in Holtzman et al. 1995a) to and
$E(B-V)$ of 0.54 and thus an distance modulus of 14.29 which
corresponds to a physical distance of 7.2 kpc.  The derived distance,
is compatible with that found by  Richtler et al. (1998) who used the
magnitude of the horizontal branch to determine the distance to NGC
65628. With $m_v(HB)=17.21\pm 0.05$ they found that the distance to
NGC 6528 is  between 6.0 and 8.9 kpc depending on the exact value for
reddening and metallicity as well as the relation between  magnitude
for the horizontal branch  and metallicity (see their Table 10). Since
we have a better handle on the metallicity we are able to be more sure
about the distance and reddening. Note that the reddening that  we
derive here is a ``minimum'' reddening in the sense that we  have
dereddened the stars on WF3 relative to those on WF2 according to the
differential reddenings found.

\section{Discussion and summary}

Colour-magnitude diagrams of  the metal-rich globular cluster NGC 6528
are notoriously difficult to analyze. This is due to the fact the
cluster is situated in the Galactic bulge and thus the fields stars
belonging to the bulge have the same magnitudes as the stars in NGC
6528.

Using two epochs of observations with HST/WFPC2 we obtain the stellar
proper motions for all stars in the field. The proper motion are  used
to separate the bulge from the cluster stars. The stellar sequences in
the resulting colour-magnitude diagram are better defined than in any
previously published colour-magnitude diagram.

Using $\alpha$-enhanced stellar isochrones we find NGC 6528 to have a
probable age of $11\pm 2$ Gyrs, this is the first attempt to establish
the absolute age of NGC 6528. Previous studies have only compared the
fiducial ridge line for the cluster to that of other globular clusters
of similar metallicities.  Mainly the comparisons have been with
regards to NGC 6553 and 47 Tuc.  With the new metallicity
determinations for individual stars in both NGC 6553 and NGC 6528 it
is now clear that 47 Tuc  (at $-0.71$ dex) has a  significantly lower
metallicity than NGC 6528 and NGC 6553 and is thus not a suitable
comparison cluster as regards differential age determinations (see
e.g. Stetson et al. 1996 and references therein).

The distance modulus obtained when fitting the stellar isochrones
yields a distance to NGC 6528 of 7.2 kpc, fully compatible with
previous derivations.

A comparison with the fiducial line for NGC 6553 confirms results in
earlier studies, e.g. Ortolani et al. (1995), that the two clusters
indeed have very similar ages.

The stellar proper motion also provide velocity dispersions for both
the cluster and field stars. The velocity dispersion of the cluster is
most likely dominated by measurement errors. The bulge dispersion can
thus be found by deconvolution. The resulting dispersions are
consistent with what has previously (Spaenhauer et al. 1992) been
found for Bulge giants. Moreover, combining our results with those by
Zoccali et al.  (2001) we are able to confirm the difference in
$\sigma_{\rm l}$ at two  positions in the bulge as predicted by the model in
Zhao (1996).

To our knowledge, our and Zoccali et al.'s study are the first to
address the proper motions amongst bulge horizontal branch and fainter
stars.

Finally we derive the mean proper motion of NGC 6528 relative to  the
Galactic bulge and also the space velocities. We find  $<\mu_{\rm l}>=\sim
+0.006$ and $<\mu_{\rm b}>=\sim +0.044$ arcsec per century and
($\Pi$,$\Theta$,$W$)=(-142, 303, 4)  km s$^{-1}$.

\acknowledgement{We thank the Royal Swedish Academy of 
Sciences for a collaborative grant that enabled
visits to Cambridge, for SF, and to Lund, for RAJ }

\begin{appendix}

\section{Stellar proper motions, coordinates and magnitudes}
\label{app:prop}

\begin{table*}
\caption[]{Stellar proper motions, both in galactic coordinates as 
well as in position on WF2-chip. The proper motions in galactic
coordinates are given in arcsec per century. Magnitudes and errors on
magnitudes are also given. The table contains data for all
stars brighter than $V_{\rm 555}=19$ and which did not 
have fitting errors. The position on the WF chip are 
for the new images. The full table (containing 212 entries) is 
available electronically from CDS }
\begin{tabular}{rrrrrrrrrrrrrrr}
\hline\noalign{\smallskip}
$\mu_{\rm l}$&$\mu_{\rm b}$& $l$&$b$&$\Delta x$&$\Delta y$& $x$&$y$&$V_{\rm 555}$&$\sigma_{V555}$&$I_{\rm 814}$ &$\sigma_{I814}$ \\
\noalign{\smallskip}
\hline\noalign{\smallskip}
--0.134 & --0.326 & 1.142 & --4.155 & --0.350 & --0.038& 614.45 & 42.80 & 17.18 & 0.001 & 15.33 & 0.003\\
  0.162 &   0.038 & 1.139 & --4.169 &   0.082 &   0.145& 120.38 & 48.03 & 17.15 & 0.002 & 15.41 & 0.003\\
--0.265 & --0.065 & 1.140 & --4.164 & --0.136 & --0.237& 296.42 & 48.49 & 17.35 & 0.002 & 16.00 & 0.004\\
  0.069 &   0.023 & 1.139 & --4.167 &   0.041 &   0.060& 179.20 & 54.49 & 16.95 & 0.001 & 15.30 & 0.003\\
... & ... & ... & ... & ... & ... & ... & ... & ... & ... & ... & ... \\
... & ... & ... & ... & ... & ... & ... & ... & ... & ... & ... & ... \\
... & ... & ... & ... & ... & ... & ... & ... & ... & ... & ... & ... \\
\noalign{\smallskip}
\hline
\end{tabular}
\label{propwf2.tab}
\end{table*}

\begin{table*}
\caption[]{Stellar proper motions, both in galactic coordinates as 
well as in position on WF3-chip. The proper motions in galactic
coordinates are given in arcsec per century. Magnitudes and errors on
magnitudes are also given. The table contains data for all
stars brighter than $V_{\rm 555}=19$ and which did not 
have fitting errors. The position on the WF chip are 
for the new images. The full table (containing 214 entries) is 
available electronically from CDS }
\begin{tabular}{rrrrrrrrrrrrrrr}
\hline\noalign{\smallskip}
$\mu_{\rm l}$&$\mu_{\rm b}$& $l$&$b$&$\Delta x$&$\Delta y$& $x$&$y$&$V_{\rm 555}$&$\sigma_{V555}$&$I_{\rm 814}$ &$\sigma_{I814}$ \\
\noalign{\smallskip}
\hline\noalign{\smallskip}
--0.207 &--0.077 & 1.146 &--4.173 &--0.177 &0.132 & 354.99 & 61.06 & 18.64 & 0.005 & 17.14 & 0.009\\
  0.102 &--0.039 & 1.145 &--4.173 &  0.109 &0.009 & 327.54 & 74.75 & 18.80 & 0.004 & 17.15 & 0.007\\
--0.080 &--0.076 & 1.145 &--4.173 &--0.055 &0.095 & 340.09 & 76.76 & 16.99 & 0.001 & 15.27 & 0.003\\
  0.024 &--0.101 & 1.147 &--4.175 &  0.051 &0.090 & 398.59 &101.47 & 17.00 & 0.001 & 15.29 & 0.003\\
... & ... & ... & ... & ... & ... & ... & ... & ... & ... & ... & ... \\
... & ... & ... & ... & ... & ... & ... & ... & ... & ... & ... & ... \\
... & ... & ... & ... & ... & ... & ... & ... & ... & ... & ... & ... \\
\noalign{\smallskip}
\hline
\end{tabular}
\label{propwf3.tab}
\end{table*}

\begin{table*}
\caption[]{Stellar proper motions, both in galactic coordinates as 
well as in position on WF4-chip. The proper motions in galactic
coordinates are given in arcsec per century. Magnitudes and errors on
magnitudes are also given. The table contains data for all
stars brighter than $V_{\rm 555}=19$ and which did not 
have fitting errors. The position on the WF chip are 
for the new images. The full table (containing 227 entries) is 
available electronically from CDS }
\begin{tabular}{rrrrrrrrrrrrrrr}
\hline\noalign{\smallskip}
$\mu_{\rm l}$&$\mu_{\rm b}$& $l$&$b$&$\Delta x$&$\Delta y$& $x$&$y$&$V_{\rm 555}$&$\sigma_{V555}$&$I_{\rm 814}$ &$\sigma_{I814}$ \\
\noalign{\smallskip}
\hline\noalign{\smallskip}
  0.528 & 0.055 & 1.136 &--4.175 &--0.205 &--0.490 & 217.89 & 51.06 & 17.29 & 0.005 & 15.51 & 0.006\\
--0.032 & 0.045 & 1.136 &--4.175 &--0.034 &  0.044 & 216.26 & 55.69 & 17.23 & 0.005 & 15.46 & 0.005\\
--0.186 & 0.115 & 1.135 &--4.177 &--0.056 &  0.211 & 292.08 & 55.84 & 18.81 & 0.009 & 17.25 & 0.017\\
  0.868&--0.076 & 1.137 &--4.172 &--0.178 &--0.853 &  88.79 & 57.58 & 17.35 & 0.004 & 15.52 & 0.005\\
... & ... & ... & ... & ... & ... & ... & ... & ... & ... & ... & ... \\
... & ... & ... & ... & ... & ... & ... & ... & ... & ... & ... & ... \\
... & ... & ... & ... & ... & ... & ... & ... & ... & ... & ... & ... \\
\noalign{\smallskip}
\hline
\end{tabular}
\label{propwf4.tab}
\end{table*}

\end{appendix}

\end{document}